\documentclass[12pt,preprint,aps,prd,nofootinbib,superscriptaddress, tightenlines]{revtex4}

\usepackage[OT1]{fontenc}
\usepackage[utf8]{inputenc}
\usepackage{epsfig}
\usepackage{epstopdf}
\usepackage[dvipsnames]{xcolor}
\usepackage{graphics}
\usepackage{lscape}
\usepackage{rotfloat}
\usepackage{rotating}
\usepackage{amsmath}
\usepackage{amssymb}
\usepackage{graphicx}                                                                        
\usepackage{flexisym}
\usepackage{breqn}
\usepackage{mathtools}
\usepackage{slashed}
\usepackage{hyperref}
\usepackage{enumitem}

\newcommand{\bdm}{\begin{dmath}}
\newcommand{\edm}{\end{dmath}}
\newcommand{\bdms}{\begin{dmath*}}
\newcommand{\edms}{\end{dmath*}}
\newcommand{\bdg}{\begin{dgroup*}}
\newcommand{\edg}{\end{dgroup*}}  
\newcommand{\be}{\begin{equation}}
\newcommand{\ee}{\end{equation}}
\newcommand{\bea}{\begin{eqnarray}} 
\newcommand{\eea}{\end{eqnarray}}
\def\openone{\leavevmode\hbox{\small1\kern-3.3pt\normalsize1}}
\newcommand{\imag}{\mbox{i}}
\newcommand{\MSbar}{\overline{\textrm{MS}}} 

\definecolor{mygreen}{RGB}{30,160,10}

\definecolor{myorange}{RGB}{255,130,0}

\begin{document}
\title{Renormalization of Wilson-line operators in the presence of nonzero quark masses}
\author{Gregoris Spanoudes} 
\email{spanoudes.gregoris@ucy.ac.cy}
\affiliation{$\it{Department \ of \ Physics, \ University \ of \ Cyprus, \ POB \ 20537, \ 1678, \ Nicosia, \ Cyprus}$ \vspace{0.5cm}}
\author{Haralambos Panagopoulos} 
\email{haris@ucy.ac.cy}
\affiliation{$\it{Department \ of \ Physics, \ University \ of \ Cyprus, \ POB \ 20537, \ 1678, \ Nicosia, \ Cyprus}$ \vspace{0.5cm}}
\begin{abstract} 
\vspace{0.5cm} In this paper, we examine the effect of nonzero quark masses on the renormalization of gauge-invariant nonlocal quark bilinear operators, including a finite-length Wilson line (called Wilson-line operators). These operators are relevant to the definition of parton quasi-distribution functions, the calculation on the lattice of which allows the direct nonperturbative study of the corresponding physical parton distribution functions. We present our perturbative calculations of the bare Green's functions, the renormalization factors in RI$'$ and $\MSbar$ schemes, as well as the conversion factors of these operators between the two renormalization schemes. Our computations have been performed in dimensional regularization at one-loop level, using massive quarks. The conversion factors can be used to convert the corresponding lattice nonperturbative results to the $\MSbar$ scheme, which is the most widely used renormalization scheme for the analysis of experimental data in high-energy physics. Also, our study is relevant for disentangling the additional operator mixing which occurs in the presence of nonzero quark masses, both on the lattice and in dimensional regularization.
\end{abstract}
\hspace{-5mm}\begin{minipage}{\textwidth}
\maketitle
\end{minipage}
\renewcommand{\thefootnote}{\alph{footnote}}
\footnotetext{Electronic addresses: ${}^{a}$ spanoudes.gregoris@ucy.ac.cy, \ ${}^{b}$ haris@ucy.ac.cy}
\renewcommand{\thefootnote}{\arabic{footnote}}
\section{Introduction}
Parton quasi-distribution functions (quasi-PDFs) are nowadays widely employed in the nonperturbative study of hadron structure in lattice QCD. They are directly related to the matrix elements of gauge-invariant nonlocal fermion bilinear operators, including a finite-length Wilson line, which are called Wilson-line operators. These functions were first introduced by X. Ji \cite{Ji:2013dva, Ji:2014gla} in order to obtain nonperturbative results for the physical light-cone parton distribution functions (PDFs) on the Euclidean lattice. PDFs are an essential tool for studying the quark and gluon structure of hadrons, as they describe the distributions of momentum and spin of constituent partons (quarks, antiquarks, and gluons) inside a hadron, in the infinite momentum frame. With the use of Large Momentum Effective Field Theory (LaMET), quasi-PDFs can be related to the physical PDFs at large momenta \cite{Ji:2013dva, Xiong:2013bka}, through a matching procedure.

\bigskip 

So far, quasi-PDFs have been studied from many points of view. Several aspects are being investigated both perturbatively and nonperturbatively, using various techniques. Exploratory lattice simulations \cite{Lin:2014zya, Alexandrou:2014pna, Gamberg:2014zwa, Alexandrou:2015rja, Gamberg:2015opc, Chen:2016utp, Bacchetta:2016zjm, Alexandrou:2016bud, Alexandrou:2016jqi, Alexandrou:2016eyt}, as well as perturbative one-loop calculations \cite{Carlson:2017gpk, Xiong:2017jtn, Constantinou:2017sej} of quasi-PDFs for the unpolarized, helicity, and transversity cases, have been performed, giving promising results. Furthermore, perturbative calculations of the matching between quasi-PDFs and physical PDFs have been implemented in Refs.\cite{Xiong:2013bka, Ma:2014jla, Ma:2014jga, Li:2016amo, Chen:2016fxx, Wang:2017qyg, Izubuchi:2018srq}; a discussion about subtleties on the continuation of PDFs to the Euclidean region can be found in Refs.\cite{Carlson:2017gpk, Briceno:2017cpo, Rossi:2017muf, Ji:2017rah}. The quasi-PDF framework is also applied to transverse momentum-dependent distributions (TMDs) \cite{Ji:2014hxa, Engelhardt:2015xja, Radyushkin:2016hsy, Radyushkin:2017ffo, Yoon:2017qzo, Broniowski:2017gfp, Ji:2018hvs}, generalized parton distributions (GPDs) \cite{Ji:2015qla, Xiong:2015nua}, hadronic light-cone distribution amplitudes (DA) \cite{Jia:2015pxx, Radyushkin:2017gjd, Zhang:2017bzy, Broniowski:2017wbr, Chen:2017gck}, and proton spin structure \cite{Ji:2014lra}. An overview of recent progress in the study of quasi-PDFs can be found in Ref. \cite{Lin:2017snn}.

\bigskip

An important issue, which needs to be addressed in order to obtain meaningful results from lattice investigations, is the renormalization of quasi-PDFs in a fully nonperturbative manner. Using a continuum regularization, Refs.\cite{Ji:2015jwa, Ishikawa:2017faj} address the renormalizability of quasi-PDFs to higher orders in perturbation theory; some related early seminal work regarding the renormalization of Wilson-loop operators can be found in Refs.\cite{Dotsenko:1979wb,Brandt:1981kf}. A perturbative one-loop calculation \cite{Constantinou:2017sej, Constantinou:2017fiy} of the matrix elements of the Wilson-line operators on the lattice has shown two nontrivial features of these operators: linear divergences (similar to those found in the continuum \cite{Dotsenko:1979wb}), in addition to the logarithmic divergences, and mixing among certain pairs of the original operators under renormalization. Studies for the elimination of the linear divergences have been made using various methods, such as the static quark potential \cite{Ma:2014jla, Ishikawa:2016znu, Ishikawa:2017jtf}, the gradient flow \cite{Monahan:2015lha, Monahan:2016bvm, Monahan:2017hpu}, the nonperturbative bare matrix elements of the Wilson-line operators \cite{Constantinou:2017sej, Constantinou:2017fiy}, and the auxiliary field formalism \cite{Dorn:1986dt}, \cite{Chen:2016fxx, Ji:2017oey, Green:2017xeu, Wang:2017eel}. A complete nonperturbative renormalization prescription, which relies on nonperturbative matrix elements of Wilson-line operators, is described in Ref.\cite{Alexandrou:2017huk}; results from recent lattice simulations by ETMC, employing this renormalization prescription, are presented in Refs.\cite{Alexandrou:2017qpu, Alexandrou:2017dzj}. A similar renormalization prescription is described in Refs.\cite{Chen:2017mzz, Lin:2017ani, Stewart:2017tvs}. Furthermore, improved lattice versions of Wilson-line operators of order $\mathcal{O} (a^1)$ are presented in Ref.\cite{Chen:2017mie}. In addition, alternative approaches for extracting physical PDFs on the lattice are currently investigated, e.g., Ioffe-time distributions (called pseudo-PDFs) \cite{Radyushkin:2017cyf, Orginos:2017kos, Radyushkin:2018cvn, Zhang:2018ggy}, Compton amplitudes utilizing the operator product expansion \cite{Chambers:2017dov}, ``lattice cross sections'' \cite{Ma:2014jla, Ma:2017pxb}, and Gaussian-weighted quasi-PDFs \cite{Chen:2017lnm}.   

\bigskip

To date, all lattice studies of the renormalization of Wilson-line operators have only considered massless fermions, expecting that the presence of quark masses can cause only imperceptible changes; this is indeed a reasonable assumption for light quarks. However, for heavy quarks this statement does not hold. In addition, simulations cannot be performed exactly at zero renormalized mass. One could, of course, adopt a zero-mass renormalization scheme even for heavy quarks, but such a scheme is less direct and entails additional complications. Thus, it would be useful to investigate the significance of finite quark masses on the renormalization of Wilson-line operators. This is the goal of our present study. 

\bigskip

In this work, we calculate the conversion factors from RI$'$ to the $\MSbar$ scheme, in dimensional regularization (DR) at one-loop level for massive quarks. The conversion factors can be combined with the regularization independent (RI$'$)-renormalization factors of the operators, computed in lattice simulations, in order to calculate the nonperturbative renormalization of these operators in $\MSbar$. Nonperturbative evaluations of the renormalization factors cannot be obtained directly in the $\MSbar$ scheme, since the definition of $\MSbar$ is perturbative in nature; most naturally, one calculates them in a RI$'$-like scheme, and then introduces the corresponding conversion factors between RI$'$ and the $\MSbar$ scheme, which rely necessarily on perturbation theory. Given that the conversion between the two renormalization schemes does not depend on regularization, it is more convenient to evaluate it in DR. Thus, the perturbative calculation of conversion factors is an essential ingredient for a complete study of quasi-PDFs. This work is a continuation to a previous paper \cite{Constantinou:2017sej}, in which, among other results, one-loop conversion factors of Wilson-line operators are presented for the case of massless quarks.  

\bigskip

In studying composite operators, one issue which must be carefully addressed is that of possible mixing with other similar operators. Many possibilities are potentially present for the nonlocal operators which we study: 
\begin{enumerate}[label=(\Alph*)]
\item Operators involving alternative paths for the Wilson line joining the quark pair will not mix among themselves, as demonstrated in Ref.\cite{Dorn:1986dt} (and also in Refs.\cite{Dotsenko:1979wb, Brandt:1981kf} for the case of closed Wilson loops). This property is related to translational invariance and is similar to the lack of mixing between a local composite operator $\mathcal{O}(x)$ with $\mathcal{O} (y)$. Given that a discrete version of translational invariance is preserved on the lattice, nonlocal operators involving different paths should not mix also on the lattice. 
\item Operators involving only gluons will also not mix. This can be seen, e.g., via the auxiliary field approach (e.g., Ref.\cite{Dorn:1986dt}); as a specific case, the operator of Eq. \eqref{O_Gamma_def} cannot mix with an operator containing the gluon field strength tensor in lieu of the quark fields (joined by a Wilson line in the adjoint representation), since this operator is higher dimensional. 
\item There may also be mixing among operators with different flavor content in a RI$'$ scheme, depending on the scheme's precise definition. However, the mixing is expected to be at most finite and thus not present in the $\MSbar$ scheme; by comparing to the massless case, in which exact flavor symmetry allows no such mixing, the difference between the massive and massless case will bear no superficial divergences, since the latter are UV regulated by the masses. The auxiliary field approach, by involving only composite operators in the (anti-)fundamental representation of the flavor group, shows that no flavor mixing needs to be introduced.
\end{enumerate}

\bigskip

Even in the absence of quark masses, bare Green's functions of Wilson-line operators may contain finite, regulator-dependent contributions which cannot be removed by a simple multiplicative renormalization; as a consequence, an appropriate (i.e., regularization-independent) choice of renormalization prescription for RI' necessitates the introduction of mixing matrices for certain pairs of operators \cite{Constantinou:2017sej}, both in the continuum and on the lattice. The results of the present work demonstrate that the presence of quark masses affects the observed operator-mixing pairs, due to the chiral-symmetry breaking of mass terms in the fermion action. Compared to the massless case on the lattice \cite{Constantinou:2017sej}, the mixing pairs remain the same for operators with equal masses of external quark fields, i.e., $(\openone, \gamma_1)$, $(\gamma_5 \gamma_2, \gamma_3 \gamma_4)$, $(\gamma_5 \gamma_3, \gamma_4 \gamma_2)$, and $(\gamma_5 \gamma_4, \gamma_2 \gamma_3)$, where by convention 1 is the direction of the straight Wilson line and 2, 3, and 4 are directions perpendicular to it. However, for operators with different masses of external quark fields, flavor-symmetry breaking leads to additional mixing pairs: $(\gamma_5, \gamma_5 \gamma_1)$, $(\gamma_2,\gamma_1 \gamma_2)$, $(\gamma_3,\gamma_1 \gamma_3)$, and $(\gamma_4,\gamma_1 \gamma_4)$.  As a consequence, the conversion factors are generally nondiagonal $2 \times 2$ matrices. This is relevant for disentangling the observed operator mixing on the lattice. Also, comparing the massive and the massless cases, the effect of finite mass on the renormalization of Wilson-line operators becomes significant for strange quarks, as well as for heavier quarks. These are features of massive quasi-PDFs, which must be taken into account in their future nonperturbative study.  

\bigskip

The outline of this paper is as follows. In Sec. II we provide the theoretical setup related to the definition of the operators which we study, along with the necessary prescription of the renormalization schemes. Sec. III contains our results for the bare Green's functions in DR, the renormalization factors, as well as the conversion factors of these operators between the renormalization schemes. In Sec. IV, we present several graphs of the conversion factor matrix elements for certain values of free parameters. Finally, in Sec. V, we conclude with possible future extensions of our work. 

\bigskip

We have also included two Appendices. Appendix A contains a discussion on technical aspects, such as the methods that we used to calculate the momentum-loop integrals, as well as the limits of vanishing regulator and/or masses. A table of Feynman parameter integrals, which appear in the expressions of our results, is relegated to Appendix B.

\section{Theoretical Setup}
\subsection{Definition of Wilson-line operators}
The Wilson-line operators are defined by a quark and an antiquark field in two different positions, a product of Dirac gamma matrices and a path-ordered exponential of the gauge field (called Wilson line), which joins the fermion fields together, in order to ensure gauge invariance. For simplicity, we choose the Wilson line to be a straight path of length z in the $\mu$ direction\footnote{For the sake of definiteness, we will often choose $\mu = 1$ in the sequel.}; thus, the operators have the form:
\begin{gather}
\mathcal{O}_\Gamma = \bar\psi(x) \Gamma \mathcal{P} \Bigg\{ \exp \left(\imag g \int_0^z d \zeta A_\mu (x + \zeta \hat{\mu})\right) \Bigg\} \psi(x + z \hat{\mu}), 
\label{O_Gamma_def}
\end{gather}
where $\Gamma = \openone,\,\gamma_5,\,\gamma_{\mu},\,\gamma_{\nu},\,\gamma_5\,\gamma_{\mu},\,\gamma_5\,\gamma_{\nu},\, \gamma_{\mu}\,\gamma_{\nu},\, \gamma_{\nu}\,\gamma_{\rho}, \ \mu \neq \nu \neq \rho \neq \mu$, and $z$ is the length of the Wilson line; $\gamma_5 = \gamma_1 \gamma_2 \gamma_3 \gamma_4$\,. The quark and antiquark fields may have different flavors: $\psi_f$ and $\bar{\psi}_{f'}$; flavor indices will be implicit in what follows. Operators with $\Gamma =$ ($\gamma_\mu$ or $\gamma_\nu$), ($\gamma_5 \gamma_\mu$ or $\gamma_5 \gamma_\nu$), ($\gamma_\mu \gamma_\nu$ or $\gamma_\nu \gamma_\rho$) correspond to the three types of PDFs: unpolarized, helicity, and transversity, respectively. 

\subsection{Definition of renormalization schemes}
Taking into account the presence of nonzero fermion masses in our calculations, we adopt mass-dependent prescriptions for the renormalization of Wilson-line operators. We define the renormalization factors which relate the bare $\mathcal{O}_\Gamma$ with the renormalized operators $\mathcal{O}^R_{\Gamma}$ via\footnote{All renormalization factors, generically labeled $Z$, depend on the regularization $X$ ($X$ = DR, LR, etc.) and on the renormalization scheme $Y$ ($Y$ = $\MSbar$, RI$'$, etc.) and should thus properly be denoted as $Z^{X,Y}$, unless this is clear from the context.}
\be 
\mathcal{O}^R_{\Gamma} = Z_\Gamma^{-1} \mathcal{O}_\Gamma. 
\label{ZGamma}
\ee
[In the presence of operator mixing, this relationship is appropriately generalized; see Eq. \eqref{EqOGamma}]. The corresponding renormalized one-particle irreducible (1-PI) amputated Green's functions of Wilson-line operators $\Lambda^R_{\Gamma} = {\langle\psi^R\,{\mathcal O}^R_{\Gamma}\,\bar \psi^R \rangle}_{\rm amp}$ are given by
\be
\Lambda^R_\Gamma = Z_{\psi_f}^{1/2} Z_{\psi_{f'}}^{1/2} \, Z_{\Gamma}^{-1} \Lambda_{\Gamma}, 
\label{LambdaGamma}
\ee
where $\Lambda_{\Gamma} = {\langle\psi\,{\mathcal O}_{\Gamma}\,\bar \psi \rangle}_{\rm amp}$ are the bare amputated Green's functions of the operators and $Z_{\psi_f}$ is the renormalization factor of the fermion field with flavor $f$, defined by $\psi^R_f  = Z^{-1/2}_{\psi_f} \psi_f$ [$\psi_f (\psi_f^R)$ is the bare (renormalized) fermion field]. In the massive case, renormalization factors of the fermion and antifermion fields appearing in bilinear operators of different flavor content may differ among themselves, as the fields have generally different masses.

\subsubsection{\underline{\textbf{Renormalization conditions for fermion fields and masses}}}
At this point, we provide the conditions for the mass-dependent renormalization of fermion fields, as well as the multiplicative renormalization of fermion masses: $m^R = Z_m^{-1} m^B$ [$m^B$ ($m^R$) are the bare (renormalized) masses for each flavor]; the latter is not involved in our calculations, but we include it for completeness. 

\bigskip

In $\MSbar$, renormalization factors $Z_\psi$ of the fermion field and $Z_m$ of the fermion mass must contain, beyond tree level, only negative powers of $\varepsilon$ (the regulator in DR in D dimensions,
$D \equiv 4 - 2 \varepsilon$); their values are fixed by the requirement that the renormalized fermion self-energy be a finite function of the renormalized parameters $m^\MSbar$ and $g^\MSbar$ ($g = {\mu}^\varepsilon Z_g g^\MSbar$; $\mu$ is a dimensionful scale): 
\be
\langle\psi^\MSbar \bar{\psi}^\MSbar\rangle = \lim_{\varepsilon \rightarrow 0} \Bigg( Z_\psi^{-1} \langle\psi \bar{\psi}\rangle \Bigg\vert_{
\begin{smallmatrix}
g = {\mu}^\varepsilon Z_g g^{\MSbar} \\
m^B \rightarrow m^\MSbar 
\end{smallmatrix}} \Bigg).
\ee

\bigskip

In RI$'$, convenient conditions for the fermion field of a given flavor and the corresponding mass are
\be
Z_\psi^{X, RI'} \ \textrm{tr} \Big(-\imag \slashed{q}\langle\psi \bar{\psi}\rangle^{-1}\Big) \Big|_{q_\nu = \bar{q}_\nu} = 4 N_c \ \bar{q}^2,
\ee
\be
Z_\psi^{X, RI'} \ \textrm{tr} \Big(\openone \langle\psi \bar{\psi}\rangle^{-1}\Big) \Big|_{q_\nu = \bar{q}_\nu} = 4 N_c \ m^{\text{RI}'} = 4 N_c \ {(Z_m^{X, RI'})}^{-1} m^B,
\ee
where $\bar{q}_\nu$ is the RI$'$ renormalization scale 4-vector, $m^{\text{RI}'}$ is the RI$'$-renormalized fermion mass, $N_c$ is the number of colors, and the symbol $X$ can be any regularization, such as DR or lattice. These conditions are appropriate for lattice regularizations which do not break chiral symmetry, so the Lagrangian mass $m_0$ coincides with the bare mass $m^B$, e.g., staggered/overlap/domain wall fermions. For regularizations which break chiral symmetry, such as Wilson/clover fermions, a critical mass $m_c$ is induced; one must first find the value of $m_c$ by a calibration in which one requires that the renormalized mass for a ``benchmark'' meson attains a desired value, e.g., zero pion mass, and then set $m^B = m_0 - m_c$. 

\subsubsection{\underline{\textbf{Renormalization conditions for Wilson-line operators}}}
As is standard practice, we will derive the factors $Z_\Gamma$ by imposing appropriate normalization conditions on the quark-antiquark Green's functions of ${\cal O}_\Gamma$.

\bigskip

In the spirit of $\MSbar$, $Z_\Gamma^{DR, \MSbar}$ contains, beyond tree level, only negative powers of $\varepsilon$. Here, we note that the leading poles in n-loop diagrams of bare Green's functions, $\mathcal{O} (1/\varepsilon^n)$ ($n \in \mathbb{Z}^+$), are multiples of the corresponding tree-level values and thus do not lead to any mixing. Subleading poles will not lead to divergent mixing coefficients, as is implicit in the renormalizability proofs of Refs.\cite{Dotsenko:1979wb, Brandt:1981kf, Dorn:1986dt}. So, in the $\MSbar$ scheme, we can use the standard definitions of renormalization factors, as in Eq. \eqref{ZGamma}.

\bigskip

In RI$'$, things are more complicated. There is, \textit{a priori}, wide flexibility in defining RI$'$-like normalization conditions for Green's functions. Given that no mixing is encountered in $\MSbar$ renormalization and given that any other scheme can only differ from $\MSbar$ by finite factors, one might \textit{a priori} expect to be able to adopt a deceptively simple prescription, in which RI$'$-renormalized operators are simply multiples of their bare counterparts, satisfying a standard normalization condition:
\be 
{\rm Tr}\Big[\Lambda_{\Gamma}^{{\rm RI}'}  (\Lambda_{\Gamma}^{\rm tree})^\dagger\Big]_{q_\nu = {\bar q}_\nu} \! \! {=} \ {\rm Tr}\Big[\Lambda_{\Gamma}^{\rm tree} (\Lambda_{\Gamma}^{\rm tree})^\dagger\Big] = 4 N_c,
\label{RC_operators} 
\ee
where $\Lambda_{\Gamma}^{\text{tree}} = \Gamma \exp (\imag q_\mu z)$ is the tree-level value of the Green's function of operator ${\cal O}_{\Gamma}$ and $\Lambda_\Gamma^{\text{RI}'}$ is defined through Eqs. \eqref{LambdaGamma} and \eqref{ZGamma}. There is, however, a fundamental problem with such a prescription: the renormalized Green's function resulting from Eq. \eqref{RC_operators} will depend on the regulator which was used in order to compute it (and, thus, it will not be regularization independent, as the name RI suggests). As was pointed out in Ref. \cite{Constantinou:2017sej}, bare Green's functions of ${\cal O}_\Gamma$, computed on the lattice, contain additional contributions proportional to the tree-level Green's function of ${\cal O}_{\Gamma'}$, where $\Gamma' = \Gamma\gamma_\mu + \gamma_\mu\Gamma$ (whenever the latter differs from zero). Such contributions will not be eliminated by applying the renormalization prescription of Eq. \eqref{RC_operators}, thus leading to renormalized Green's functions which differ from those obtained in DR. It should be pointed out that, in all cases, the renormalized functions will contain a number of tensorial structures, the elimination of which may be possible at best only at a given value of the renormalization scale. However, the main concern here is not the elimination of mixing contributions, desirable as this might be; what is more important is to establish a RI$'$ scheme which is indeed regularization independent, so that nonperturbative estimates of renormalization factors can be converted to the $\MSbar$ scheme using conversion factors which are regulator independent.

\bigskip

Given the preferred direction $\mu$ of the Wilson-line operator, there is a residual rotational (or hypercubic, on the lattice) symmetry with respect to the three remaining transverse directions, including also reflections. As a consequence, given an appropriate choice of a renormalization scheme, no mixing needs to occur among operators which do not transform in the same way under this residual symmetry. In particular, mixing can occur only among pairs of operators $(\mathcal{O}_\Gamma, \mathcal{O}_{\Gamma \gamma_\mu})$.

\bigskip

Denoting generically the two operators in such a pair by $(\mathcal{O}_{\Gamma_1}, \mathcal{O}_{\Gamma_2})$, the corresponding renormalization factors will be $2 \times 2$ mixing matrices:
\be 
\mathcal{O}^{RI'}_{\Gamma_i} = \sum_{j= 1}^2 \Big[{(Z_{\Gamma_1, \Gamma_2}^{X, RI'})}^{-1}\Big]_{ij} \mathcal{O}_{\Gamma_j}, \quad (i = 1,2).
\label{EqOGamma}
\ee
More precisely, the mixing pairs $(\mathcal{O}_{\Gamma_1}, \mathcal{O}_{\Gamma_2})$ are $(\openone, \gamma_1)$, $(\gamma_5, \gamma_5 \gamma_1)$, $(\gamma_2,\gamma_1 \gamma_2)$, $(\gamma_3,\gamma_1 \gamma_3)$, $(\gamma_4,\gamma_1 \gamma_4)$, $(\gamma_5 \gamma_2, \gamma_3 \gamma_4)$, $(\gamma_5 \gamma_3, \gamma_4 \gamma_2)$, and $(\gamma_5 \gamma_4, \gamma_2 \gamma_3)$. Therefore, the renormalized 1-PI amputated Green's functions of Wilson-line operators have the following form:
\be 
\Lambda^{RI'}_{\Gamma_i} = \sum_{j= 1}^2 \big(Z_{\psi_f}^{X, RI'} \big)^{1/2} \ \big(Z_{\psi_{f'}}^{X, RI'} \big)^{1/2} \, \Big[{(Z_{\Gamma_1, \Gamma_2}^{X, RI'})}^{-1}\Big]_{ij} \Lambda_{\Gamma_j}.
\label{def_Lambda_RIprime}
\ee
Thus, an appropriate renormalization condition, especially for lattice simulations, is 
\be 
{\rm Tr}\Big[\Lambda_{\Gamma_i}^{{\rm RI}'}  (\Lambda_{\Gamma_j}^{\rm tree})^\dagger\Big]_{q_\nu = {\bar q}_\nu} \! \! {=} \ {\rm Tr}\Big[\Lambda_{\Gamma_i}^{\rm tree} (\Lambda_{\Gamma_j}^{\rm tree})^\dagger\Big] = 4 N_c \ \delta_{ij}. 
\label{RC_Lambda_RIprime}
\ee
Combining Eqs. \eqref{def_Lambda_RIprime} and \eqref{RC_Lambda_RIprime}, the RI$'$ condition takes the form: 
\be 
{(Z_{\Gamma_1, \Gamma_2}^{X,{\rm RI}'})}_{ij} = \frac{1}{4 N_c} \big(Z_{\psi_f}^{X, RI'} \big)^{1/2} \ \big(Z_{\psi_{f'}}^{X, RI'} \big)^{1/2} \ {\rm Tr}\Big[\Lambda_{\Gamma_i} \, (\Lambda_{\Gamma_j}^{\rm tree})^\dagger\Big]_{q_\nu = {\bar q}_\nu}.
\label{RC_Lambda_RIprime2}
\ee 
Based on the above symmetry arguments, such a RI$'$ condition will indeed be regularization independent, for all regularizations which respect the above symmetries.

\bigskip

One could of course adopt more general definitions of RI$'$, e.g., a prescription in which each of the 16 operators ${\cal O}_\Gamma$ can contain admixtures of some of the remaining operators:
\be 
\mathcal{O}^{\text{RI}'}_{\Gamma_i} = \sum_{j=1}^{16} \Big[(Z^{\text{X},\text{RI}'})^{-1}\Big]_{ij} \mathcal{O}_{\Gamma_j}, \quad (i = 1,\cdots ,16),
\ee
in such a way that the renormalized Green's functions will  satisfy a condition similar to Eq. \eqref{RC_Lambda_RIprime}, but with the indices $i,j$ ranging from 1 to 16. However, such a definition would introduce additional finite mixing, which would violate the rotational symmetry in the transverse directions, e.g., mixing among $\mathcal{O}_{\gamma_1}$ and $\mathcal{O}_{\gamma_2}$; such a violation would occur whenever the RI$'$ renormalization scale 4-vector ${\bar q}$ is chosen to lie in an oblique direction. To avoid such unnecessary mixing, it is thus natural to adopt the ``minimal" prescription of Eqs. \eqref{EqOGamma} - \eqref{RC_Lambda_RIprime2}. Since this prescription extends beyond one-loop order, it may be applied to nonperturbative evaluations of the renormalization matrices $Z^{\text{L}, \text{RI}'}$.

\bigskip

Let us note that, as it stands, Eq. \eqref{RC_Lambda_RIprime} leads to renormalization factors which depend on the individual components of $\bar{q}$, rather than just $\bar{q}^2$ and $\bar{q}_\mu$; consequently, the renormalization factors of, e.g., $\mathcal{O}_{\gamma_2}$ and $\mathcal{O}_{\gamma_3}$ will have different numerical values. One could, of course, define RI$'$ in such a way that the residual invariance is manifest; this can be seen by analogy with local operators, e.g., $\mathcal{O}_{V_i} = \bar{\psi}(x) \gamma_i  \psi(x)$, where $Z_V$ is often defined as the average over $Z_{V_i}$ ($i$ = 1,2,3,4), and, in doing so, $Z_V$ turns out to depend only on the length of the renormalization scale 4-vector. Adopting such a definition, the values of the conversion factors can be read off our bare Green's functions [see Eqs. \eqref{LambdaSd1} - \eqref{Lambdad4} below] in a rather straightforward way, and they will indeed depend only on $\bar{q}^2$ and $\bar{q}_\mu$. However, in defining the RI$'$ scheme for Wilson-line operators, we have aimed at being as general as possible and thus did not take any averages, as above, in order to accommodate possible definitions employed in nonperturbative investigations of the renormalization factors; after all, the conversion factors which we calculate must be applicable precisely to these investigations. It goes without saying that if one chooses all components of the renormalization scale 4-vector, perpendicular to the Wilson line, to vanish, then residual rotational invariance is automatically restored. 

\bigskip 

Finally, one could define RI$'$ in such a way that renormalization factors would be strictly real, e.g., by taking the absolute value of the lhs in Eq.  \eqref{RC_Lambda_RIprime}; indeed, the choice of the definition of RI$'$, leading to complex renormalization factors, is not mandatory, but it is a natural one, following the definition used in nonperturbative investigations. All these choices are related to the $\MSbar$ scheme via finite conversion factors; thus, no particular choice is dictated by the need to remove divergences, either in dimensional regularization or on the lattice.

\subsection{Conversion factors}
As a consequence of the $2 \times 2$ matrix form of the RI$'$ renormalization factors, the conversion factors between RI$'$ and $\MSbar$ schemes will also be $2 \times 2$ mixing matrices. Being regularization independent, they can be evaluated more easily in DR. They are defined as 
\be
\Big[\mathcal{C}_{\Gamma_1, \Gamma_2}^{\MSbar,{\rm RI}'}\Big]_{ij} = (Z_{\Gamma_i}^{DR,\MSbar})^{-1} \cdot \Big[Z_{\Gamma_1,\Gamma_2}^{DR,{\rm RI}'}\Big]_{ij} = \sum_{k=1}^2 \Big[(Z_{\Gamma_1, \Gamma_2}^{LR,\MSbar})^{-1}\Big]_{ik} \cdot \Big[Z_{\Gamma_1, \Gamma_2}^{LR,{\rm RI}'}\Big]_{kj}.
\label{ConvFactors}
\ee
We note in passing that the definition of the $\MSbar$ scheme depends on the prescription used for extending $\gamma_5$ to D dimensions\footnote{See, e.g., Refs. \cite{Buras:1989xd,Patel:1992vu,Larin:1993tp,Larin:1993tq,Skouroupathis:2008mf,Constantinou:2013pba} for a discussion of four relevant prescriptions and some conversion factors among them.}; this, in particular, will affect conversion factors for the pseudoscalar and axial-vector operators. However, such a dependence will only appear beyond one loop. Now, the Green's functions in the RI$'$ scheme can be directly converted to the $\MSbar$ scheme through
\begin{align}
\binom{\Lambda_{\Gamma_1}^\MSbar}{\Lambda_{\Gamma_2}^\MSbar} &= 
{\left(Z_{\psi_f}^{LR,\MSbar} \over Z_{\psi_f}^{LR,{\rm RI}'}\right)}^{1/2} {\left(Z_{\psi_{f'}}^{LR,\MSbar} \over Z_{\psi_{f'}}^{LR,{\rm RI}'}\right)}^{1/2} \, (Z_{\Gamma_1, \Gamma_2}^{LR,\MSbar})^{-1} \cdot (Z_{\Gamma_1, \Gamma_2}^{LR,{\rm RI}'}) \cdot 
\binom{\Lambda_{\Gamma_1}^{{\rm RI}'}}{\Lambda_{\Gamma_2}^{{\rm RI}'}} \nonumber \\
&=
{1 \over \big(\mathcal{C}_{\psi_f}^{\MSbar,{\rm RI}'} \big)^{1/2} \ \big(\mathcal{C}_{\psi_{f'}}^{\MSbar,{\rm RI}'} \big)^{1/2}}\, (\mathcal{C}_{\Gamma_1, \Gamma_2}^{\MSbar,{\rm RI}'})\, 
\cdot \binom{\Lambda_{\Gamma_1}^{{\rm RI}'}}{\Lambda_{\Gamma_2}^{{\rm RI}'}},
\label{convGreen}
\end{align}
where $\mathcal{C}_{\psi_f}^{\MSbar,{\rm RI}'} \equiv Z_{\psi_f}^{LR,{\rm RI}'} /Z_{\psi_f}^{LR,\MSbar} 
= Z_{\psi_f}^{DR,{\rm RI}'} /Z_{\psi_f}^{DR,\MSbar}$ is the conversion factor for a fermion field of a given flavor.

\section{Computation and Results}

In this section, we present our one-loop results for the bare Green's functions of Wilson-line operators, the renormalization factors, and the conversion factors between RI$'$ and $\MSbar$ schemes, using dimensional regularization. In this regularization, Green's functions are Laurent series in $\varepsilon$, where $\varepsilon$ is the regulator, defined by $D \equiv 4 - 2 \varepsilon$, and $D$ is the number of Euclidean spacetime dimensions, in which momentum-loop integrals are well defined. We also investigate the operator mixing.
\subsection{Bare Green's functions}
\begin{figure}[thb] 
  \centering
  \includegraphics[width=16.5cm,clip]{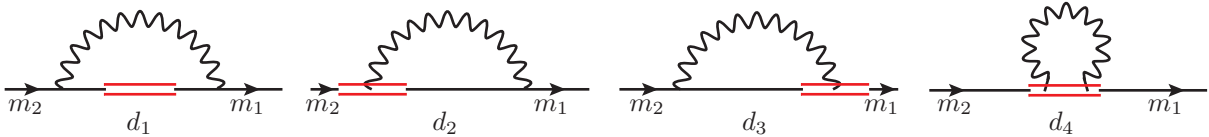}
  \caption{Feynman diagrams contributing to the one-loop calculation of the Green's functions of Wilson-line operator $\mathcal{O}_\Gamma$. The straight (wavy) lines represent fermions (gluons). The operator insertion is denoted by double straight line.}
  \label{Fig.FeynmanDiagrams}
\end{figure}

There are four one-loop Feynman diagrams corresponding to the two-point Green's functions of operators $\mathcal{O}_\Gamma$, shown in Fig. \ref{Fig.FeynmanDiagrams}. 
The last diagram ($d_4$) does not depend on the quark masses, and therefore its contribution is the same as that of the massless case. In Appendix A, we describe the method that we used to calculate the momentum-loop integrals of the above diagrams. Below, we provide our results for the bare Green's function of operators for each Feynman diagram separately. Our expressions depend on integrals of modified Bessel functions of the second kind, $K_n$, over Feynman parameters. These integrals are presented in Eqs. \eqref{f1} - \eqref{h8} of Appendix B. For the sake of brevity, we use the following notation: $f_{ij} \equiv f_i \left(q,z,m_j\right)$, \ $g_{ij} \equiv g_i \left(q,z,m_j\right)$, and $h_i \equiv h_i \left(q,z,m_1,m_2\right)$. Also, index $\mu$ is the direction parallel to the Wilson line; indices $\nu$, $\rho$, and $\sigma$ are the directions perpendicular to the Wilson line; and $\mu$, $\nu$, $\rho$, and $\sigma$ are all different among themselves. Furthermore, $\bar{\mu}$ is the $\MSbar$ renormalization scale, $\bar{\mu} \equiv \mu \ (4 \pi / e^{\gamma_E})^{1/2}$, where $\mu$ (not to be confused with the spacetime index $\mu$) appears in the renormalization of the D-dimensional coupling constant; $g = \mu^\varepsilon \ Z_g \ g^R$, and $\gamma_E$ is the Euler constant. In addition, $C_f = (N_c^2 - 1)/(2 N_c)$ is the Casimir operator, and $\beta$ is the gauge fixing parameter, defined such that $\beta = 0(1)$ corresponds to the Feynman (Landau) gauge. Finally, symbols $S$ (scalar), $P$ (pseudoscalar), $V_\mu$ (vector in the $\mu$ direction), $V_\nu$ (vector in the $\nu$ direction), $A_\mu$ (axial-vector in the $\mu$ direction), $A_\nu$ (axial-vector in the $\nu$ direction), $T_{\mu \nu}$ (tensor in the $\mu$, and $\nu$ directions), and $T_{\nu \rho}$ (tensor in the $\nu$, and $\rho$ directions) correspond to the operators ${\cal O}_\Gamma$ with $\Gamma = \openone$, \ $\gamma_5$, \ $\gamma_\mu$, \ $\gamma_\nu$, \ $\gamma_5 \gamma_\mu$, \ $\gamma_5 \gamma_\nu$, \ $\gamma_\mu \gamma_\nu$, \ $\gamma_\nu \gamma_\rho$, respectively. We note that only tree-level values for the quark masses appear in the following one-loop expressions: \\ 
\begin{align}
\Lambda_S^{\text{d}_1} = \frac{g^2 C_f}{16 \pi ^2} \Bigg\{& \Lambda_S^{\text{tree}} \Bigg[ \left(\beta - 4\right) \Big( - 4 h_1 - 2 \imag z q_\mu h_2 + \left| z \right| \left( h_4 + m_1 m_2 h_5 - q^2 h_7 \right) \Big) \nonumber \\
&\qquad \ + \beta \left(q^2 - m_1 m_2 \right) \Big[ \frac{1}{2} z^2 \left( h_2 - q^2 h_3 \right) + \left| z \right| \Big(\imag z q_\mu \left( h_5 - h_6 - h_7\right) \nonumber \\
&\qquad \qquad \qquad \qquad \qquad \ - \left( h_5 - 2 h_6 + q^2 h_8\right)\Big)\Big] \Bigg] \nonumber \\  
&+ \Lambda_S^{\text{tree}} \slashed{q} \ \imag \left(m_1+m_2\right) \Big[\beta \left(\left| z\right| \left(h_5-q^2 h_8\right)- \frac{1}{2} z^2 \left(h_2 + q^2 h_3\right)\right)-2 \left| z\right| \left(h_5-h_6\right) \Big] \nonumber \\
&+ \Lambda_{V_{\mu}}^{\text{tree}} \left(m_1+m_2\right) z \Big[(\beta +2) h_1 - \beta \Big( \left| z\right| q^2 \left(h_5-h_6-h_7\right) - \imag z q_{\mu} h_2 \Big)\Big]\Bigg\}, 
\label{LambdaSd1}
\end{align}
\vspace{0.5cm}
\be 
\Lambda_P^{\text{d}_1} = \gamma_5 \ \Lambda_S^{\text{d}_1} \{ m_2 \mapsto - m_2 \},
\ee 

\begin{align}
\Lambda_{V_\mu}^{\text{d}_1} = \frac{g^2 C_f}{16 \pi ^2} \Bigg\{& \Lambda_{V_{\mu}}^{\text{tree}} \Bigg[ -4 \left(\beta-1\right) h_1 + \left| z \right| \Big[ \left(\beta+2\right) h_4 - \left(\beta-2\right) \left(m_1 m_2 h_5 + q^2 h_7 \right) \Big] \nonumber \\
&\qquad \quad + 2 \beta z q_\mu \Big[z q_\mu h_2 - \imag \left(h_1 + h_2 \right) + \imag \left| z \right| \left( h_5-h_6-h_7\right) \Big] \nonumber \\
&\qquad \quad + \beta \left(q^2 + m_1 m_2\right) \Big[\left| z \right| \left(h_6 - q^2 h_8\right) - \frac{1}{2} z^2 \left(h_2 + q^2 h_3\right) \Big] \Bigg] \nonumber \\
&+ \left(\Lambda_{V_{\mu }}^{\text{tree}} \slashed{q} \ m_1 + \slashed{q} \Lambda_{V_{\mu }}^{\text{tree}} \ m_2 \right) \ \beta \Big[- \left| z\right|  \Big(z q_{\mu} \left(h_5-h_6-h_7\right) -\imag \left(h_6 - q^2 h_8 \right)\Big) \nonumber \\
&\qquad \qquad \qquad \qquad \qquad \qquad \quad \ + \frac{1}{2} \imag z^2 \left(h_2 - q^2 h_3 \right)\Big] \nonumber \\
&+\Lambda_S^{\text{tree}} \left( m_1 + m_2 \right) \Big[z (\beta-4) h_1 -2 \imag \left| z\right| q_\mu (\beta - 2) \left(h_5 - h_6\right)\Big] \nonumber \\
&+ \Lambda_S^{\text{tree}} \slashed{q} \Big[2 \left| z \right| q_\mu \Big( \beta \left(h_5 - h_6 + m_1 m_2 h_8\right) - 2 h_7\Big) - \beta z^2 q_\mu \left(h_2 - m_1 m_2 h_3\right) \nonumber \\
&\qquad \qquad  + 2 \imag z \left(\beta h_1 - 2 h_2\right) -\imag \beta z \left| z \right| \left(q^2 + m_1 m_2\right) \left(h_5 - h_6 - h_7\right) \Big] \Bigg\},
\end{align}
\begin{align}
\Lambda_{V_\nu}^{\text{d}_1} = \frac{g^2 C_f}{16 \pi ^2} \Bigg\{& \Lambda_{V_{\nu }}^{\text{tree}} \Bigg[ -2 \Big( 2 \left(\beta - 1\right) h_1 + \left(\beta - 2\right) \imag z q_\mu h_2 \Big) - \left( \beta - 2 \right) \left| z \right| \left(m_1 m_2 h_5 + q^2 h_7 - h_4\right) \nonumber \\
 &\qquad \ \ + \left(q^2 + m_1 m_2 \right) \beta \Big[ \left| z \right| \Big( \imag z q_\mu \left( h_5-h_6-h_7\right) + \left(h_6 - q^2 h_8 \right) \Big) \nonumber \\
 &\qquad \qquad \qquad \qquad \qquad \ \ + \frac{1}{2} z^2 \left( h_2 - q^2 h_3 \right) \Big] \Bigg] \nonumber \\
&+ \left( \Lambda_{V_{\nu}}^{\text{tree}} \slashed{q} \ m_1 + \slashed{q} \Lambda_{V_{\nu}}^{\text{tree}} \ m_2 \right) \imag \beta \Big[ \left| z\right|  \left(h_6- q^2 h_8\right) - \frac{1}{2} z^2 \left(h_2 + q^2 h_3\right) \Big] \nonumber \\
&+ \Lambda_{T_{\mu \nu }}^{\text{tree}}\left(m_1-m_2\right) \beta z \Big[-h_1 + \left| z\right| q^2 \left(h_5-h_6-h_7\right) - \imag z q_{\mu} h_2 \Big] \nonumber \\
 &+ \Lambda_{V_{\mu }}^{\text{tree}} z q_{\nu} \Big[ \beta  \Big(\imag \left| z\right|  \left(q^2-m_1 m_2\right) \left(h_5-h_6-h_7\right) +2 \left( z q_{\mu} h_2 - \imag h_1\right) \Big)-4 \imag h_2\Big] \nonumber \\
 &- \left( \Lambda_{V_{\mu }}^{\text{tree}} \slashed{q} \ m_1 + \slashed{q} \Lambda_{V_{\mu }}^{\text{tree}} \ m_2 \right) \beta z \left| z\right|  q_{\nu} \left(h_5-h_6-h_7\right) \nonumber \\
 &+ \Lambda_S^{\text{tree}} \left(m_1+m_2\right) \imag q_{\nu} \Big[-2 (\beta - 2) \left| z\right| \left(h_5-h_6\right) + \beta z^2 h_2 \Big] \nonumber \\
 &+ \Lambda_S^{\text{tree}} \slashed{q} q_{\nu} \hspace{-0.05cm}\Big[\beta \Big(2 \left| z\right| \left(h_5-h_6+ m_1 m_2 h_8 \right)-z^2 \left(h_2 - m_1 m_2 h_3\right) \Big) -4 \left| z\right| h_7 \Big]\hspace{-0.1cm}\Bigg\} \hspace{-0.29cm}
\end{align}
\be
\Lambda_{A_{\mu (\nu)}}^{\text{d}_1} = \gamma_5 \ \Lambda_{V_{\mu (\nu)}}^{\text{d}_1} \{ m_2 \mapsto - m_2 \},
\ee 
\begin{align}
\Lambda_{T_{\mu \nu}}^{\text{d}_1} = \frac{g^2 C_f}{16 \pi ^2} \Bigg\{& \Lambda_{T_{\mu  \nu}}^{\text{tree}} \beta \Bigg[ -2 \left( 2 h_1 - z^2 q^2_\mu h_2 \right) + \left| z \right| \Big[ h_4 + q^2 \Big( h_5 - h_7 + 2 \imag z q_\mu \left(h_5-h_6-h_7\right) \Big) \Big] \nonumber \\
&\qquad \quad \ -2 \imag z q_\mu (h_1 + h_2)  - \left(q^2 - m_1 m_2\right) \Big( \left| z \right| q^2 h_8 + \frac{1}{2} z^2 \left(h_2 + q^2 h_3\right) \Big) \Bigg] \nonumber \\
&+ \left(\Lambda_{T_{\mu  \nu }}^{\text{tree}} \slashed{q} \ m_1 + \slashed{q} \Lambda_{T_{\mu  \nu }}^{\text{tree}} \ m_2\right) \Bigg[\beta  \Big[\frac{1}{2} \imag z^2 \left( h_2 - q^2 h_3 \right) - \left| z\right| \Big(z q_{\mu} \left(h_5-h_6-h_7\right) \nonumber \\
&\qquad \qquad \qquad \qquad \qquad \qquad \ + \imag \left(h_5-2 h_6 + q^2 h_8 \right)\Big) \Big]+2 \imag \left| z\right| \left(h_5-h_6\right) \Bigg] \nonumber \\
&-\Lambda_{V_{\nu }}^{\text{tree}} (\beta -2) z \left(m_1-m_2\right) h_1 \nonumber \\
&+ \Lambda_{V_{\nu }}^{\text{tree}} \slashed{q} \beta \Bigg[2 \left( \left| z\right| q_{\mu} m_1 m_2 h_8 - \imag z h_1\right) + \imag z \left| z\right| \left(q^2 - m_1 m_2\right)\left(h_5-h_6-h_7\right) \nonumber \\
&\qquad \qquad \quad + z^2 q_{\mu} \left(h_2 + m_1 m_2 h_3\right) \Bigg] \nonumber \\
&- \Lambda_{V_{\mu }}^{\text{tree}} \imag \beta z^2 q_{\nu} (m_1 - m_2) h_2 + \Lambda_{V_{\mu }}^{\text{tree}} \slashed{q} \beta q_{\nu} \Big[ z^2 \left( h_2 - m_1 m_2 h_3\right) - 2 \left| z\right| m_1 m_2 h_8 \Big] \nonumber \\ 
&+ \Lambda_S^{\text{tree}} \beta z q_{\nu} \Big[2 \left(\imag h_1 - z q_{\mu} h_2 \right) -\imag \left| z\right| \left(q^2-m_1 m_2\right) \left(h_5-h_6-h_7\right) \Big] \nonumber \\
&+ \Lambda_S^{\text{tree}} \slashed{q} \beta z \left| z\right| q_{\nu} \left(m_1-m_2\right) \left(h_5-h_6-h_7\right) \Bigg\},
\end{align}

\begin{align}
\Lambda_{T_{\nu \rho}}^{\text{d}_1} = \frac{g^2 C_f}{16 \pi ^2} \Bigg\{& \Lambda_{T_{\nu  \rho}}^{\text{tree}} \beta \Bigg[ -2 \left(2 h_1 + \imag z q_\mu h_2\right) + \left| z \right| \Big( h_4 + q^2 \left( h_5 - h_7\right) \Big) \nonumber \\ 
 &\qquad \quad \ + \left(q^2 - m_1 m_2\right) \Big[ \frac{1}{2} z^2 \left( h_2 - q^2 h_3 \right) - \left| z \right| \Big( q^2 h_8 - \imag z q_\mu \left(h_5 - h_6 - h_7\right)\Big)\Big]\Bigg] \nonumber \\
&+ \left( \Lambda_{T_{\nu  \rho}}^{\text{tree}} \slashed{q} \ m_1 + \slashed{q} \Lambda_{T_{\nu  \rho}}^{\text{tree}} m_2\right) \imag  \Big[-\beta  \Big(\frac{1}{2} z^2 \left(h_2 + q^2 h_3\right) + \left| z\right| \left(h_5-2 h_6 + q^2 h_8 \right)\Big) \nonumber \\
&\qquad \qquad \qquad \qquad \qquad \qquad \ + 2 \left| z\right| \left(h_5-h_6\right)\Big] \nonumber \\
&+ \varepsilon_{\mu \nu \rho \sigma} \ \Lambda_{A_{\sigma}}^{\text{tree}} \left(m_1 + m_2 \right) \Big[-\beta  z \left| z\right| q^2  \left(h_5-h_6-h_7\right) + \left( \beta - 2 \right) z h_1 + \imag \beta z^2 q_{\mu} h_2 \Big] \nonumber \\
&+ \left( \Lambda_{T_{\mu  \nu }}^{\text{tree}} q_{\rho} - \Lambda_{T_{\mu \rho }}^{\text{tree}} q_{\nu} \right) \beta  z \Big[2 \left( \imag h_1 - z h_2 q_{\mu}\right) - \imag \left| z\right| \left(q^2+m_1 m_2\right) \left(h_5-h_6-h_7\right) \Big] \nonumber \\
&+ \Big[ \left( \Lambda_{T_{\mu  \nu }}^{\text{tree}} q_{\rho} - \Lambda_{T_{\mu \rho }}^{\text{tree}} q_{\nu} \right) \slashed{q} \ m_1 + \slashed{q} \left( \Lambda_{T_{\mu  \nu }}^{\text{tree}} q_{\rho} - \Lambda_{T_{\mu \rho }}^{\text{tree}} q_{\nu} \right) m_2 \Big] \beta z \left| z\right| \left(h_5-h_6-h_7\right) \nonumber \\
&+ \left( \Lambda_{V_{\nu }}^{\text{tree}} q_\rho - \Lambda_{V_{\rho }}^{\text{tree}} q_\nu \right) \imag \beta z^2 \left(m_1 - m_2\right) h_2 \nonumber \\
& - \left(\Lambda_{V_{\nu }}^{\text{tree}} q_{\rho} - \Lambda_{V_{\rho }}^{\text{tree}} q_{\nu} \right) \slashed{q} \beta  \Big[z^2 \left( h_2 + m_1 m_2 h_3\right) +2 \left| z\right| m_1 m_2 h_8\Big] \Bigg\},
\end{align}
\begin{align}
\hspace{-1mm}\Lambda_S^{\text{d}_2} = \frac{g^2 C_f}{16 \pi ^2} \Bigg\{& \Lambda_S^{\text{tree}} \Bigg[ (\beta -1) \Big[2 f_{11} -2 - \frac{1}{\varepsilon} - \log \left(\frac{\overline{\mu }^2}{q^2 + m_1^2}\right) + \frac{m_1^2}{q^2} \log \left(1 + \frac{q^2}{m_1^2}\right) \Big] \nonumber \\
& \qquad \ \ +\beta q^2 \Big(\imag q_{\mu } \left(g_{31}-z f_{31}\right)+ \left(q^2 + m_1^2\right) g_{41} - \left(q^2 - q_{\mu }^2\right) g_{51} \Big) -2 \imag q_{\mu} g_{21} \Bigg] \nonumber \\
&+ \Lambda_S^{\text{tree}} \slashed{q} \beta  m_1 \Big[ - q_{\mu } \left(g_{31}-z f_{31}\right)+ \imag g_{41} \left(q^2 + m_1^2\right)- \imag g_{51} \left(q^2 - q_{\mu}^2\right)\Big] \nonumber \\
&+ \Lambda_{V_{\mu }}^{\text{tree}} m_1 \left(2 g_{11}-\beta  z f_{21}\right) + \Lambda_{V_{\mu}}^{\text{tree}} \slashed{q} \imag \Big(\beta  z f_{21}-2 \left( g_{11} -  g_{21} \right) \Big) \Bigg\},
\end{align}
\be
\Lambda_\Gamma^{\text{d}_2} =  \Lambda_S^{\text{d}_2} \ \Gamma,
\ee
\begin{align}
\hspace{-1mm}\Lambda_S^{\text{d}_3} = \frac{g^2 C_f}{16 \pi ^2} \Bigg\{& \Lambda_S^{\text{tree}} \Bigg[ (\beta -1) \Big[2 f_{12} -2 - \frac{1}{\varepsilon} - \log \left(\frac{\overline{\mu }^2}{q^2 + m_2^2}\right) + \frac{m_2^2}{q^2} \log \left(1 + \frac{q^2}{m_2^2}\right) \Big] \nonumber \\
& \qquad \ \ +\beta q^2 \Big(\imag q_{\mu } \left(g_{32}-z f_{32}\right)+ \left(q^2 + m_2^2\right) g_{42} - \left(q^2 - q_{\mu }^2\right) g_{52} \Big) -2 \imag q_{\mu} g_{22} \Bigg] \nonumber \\
&+ \Lambda_S^{\text{tree}} \slashed{q} \beta  m_2 \Big[ - q_{\mu } \left(g_{32}-z f_{32}\right)+ \imag g_{42} \left(q^2 + m_2^2\right)- \imag g_{52} \left(q^2 - q_{\mu}^2\right)\Big] \nonumber \\
&+ \Lambda_{V_{\mu }}^{\text{tree}} m_2 \left(2 g_{12}-\beta  z f_{22}\right) + \slashed{q} \Lambda_{V_{\mu}}^{\text{tree}} \imag \Big(\beta  z f_{22}-2 \left( g_{12} -  g_{22} \right) \Big) \Bigg\},
\end{align}
\be
\Lambda_\Gamma^{\text{d}_3} =  \Gamma \ \Lambda_S^{\text{d}_3},  
\ee
\be
\Lambda_\Gamma^{\text{d}_4} = \frac{g^2 C_f}{16 \pi ^2} \Lambda_{\Gamma}^{\text{tree}} \Big[4 + (\beta +2) \Big( 2 \gamma_E + \frac{1}{\varepsilon} + \log \left(\frac{1}{4} z^2 \overline{\mu }^2\right)\Big)\Big].
\label{Lambdad4}
\ee

\bigskip

UV-divergent terms of order $\mathcal{O} (1/\varepsilon)$ arise from the last three diagrams. These terms are multiples of the tree-level values of Green's functions and therefore do not lead to any mixing. However, there are finite terms for each $\mathcal{O}_\Gamma$ with different Dirac structures than the original operator; some of these terms are responsible for the finite mixing which occurs in RI$'$. In particular, they lead to the expected mixing within the pairs $(\Gamma, \Gamma \gamma_\mu)$ or equivalently $(\Gamma, \gamma_\mu \Gamma)$. This is a consequence of the violation of chiral symmetry by the mass term in the fermion action, as well as the flavor-symmetry breaking when masses have different values. For the case of equal masses (no flavor-symmetry breaking) $m_1 = m_2$, the mixing pattern reduces to $(\Gamma, \frac{1}{2} \{\Gamma, \gamma_\mu \})$, which is the same as the pattern for massless quarks on the lattice. Our findings are expected to be valid also on the lattice. 

\bigskip

The one-loop Green's functions exhibit a nontrivial dependence on dimensionless quantities involving the Wilson-line length z, the external quark momentum $q$, and the quark masses $m_i$ ($i = 1, 2$): $z q_\mu$, $z m_i$. This dependence is in addition to the standard logarithmic dependence on $\bar{\mu}$: $\log (\bar{\mu}^2 / q^2)$. Also, we note that our results are not analytic functions of $z$ near $z = 0$; this was expected due to the appearance of contact terms beyond tree level. For the case $z=0$, the nonlocal operators are replaced by local massive fermion bilinear operators; their renormalization is addressed in Ref.\cite{Boyle:2016wis}, using a generalization of the RI-SMOM scheme, called RI-mSMOM. Further, the Green's functions of Feynman diagrams satisfy the following reflection relations, with respect to $z$: 
\begin{align}
\Lambda_\Gamma^{d_1} (z,m_1,m_2) &= \frac{1}{4} \text{tr} (\Gamma^2) \ {\Big[ \Lambda_\Gamma^{d_1} (-z,-m_2,-m_1)\Big]}^\dagger, \\
\Lambda_\Gamma^{d_2} (z,m) &= \frac{1}{4} \text{tr} (\Gamma^2) \ {\Big[\Lambda_\Gamma^{d_3} (-z,-m)\Big]}^\dagger, \\
\Lambda_\Gamma^{d_4} (z) &= \frac{1}{4} \text{tr} (\Gamma^2) \ {\Big[\Lambda_\Gamma^{d_4} (-z)\Big]}^\dagger.
\end{align}
[Note that $(1/4)\ \text{tr} (\Gamma^2) = \pm 1$, depending on $\Gamma$.] The total one-loop bare Green's functions of operators $\mathcal{O}_\Gamma$ are given by the sum over the contributions of the four diagrams:
\be 
\Lambda_\Gamma^{\text{1-loop}} = \sum_{i=1}^4 \Lambda_\Gamma^{d_i}.
\ee

\subsection{Renormalization factors}

\subsubsection{\underline{\textbf{Renormalization factors of fermion field and mass}}}
\begin{figure}[!thb] 
  \centering
  \includegraphics[width=4.5cm,clip]{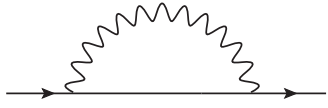}
  \caption{Feynman diagram contributing to the one-loop calculation of the fermion self-energy. The straight (wavy) lines represent fermions (gluons).}
  \label{Feynman_Diagram_Zpsi_Zm}
\end{figure}

The perturbative determination of $Z_\psi$ and $Z_m$ proceeds in textbook fashion by calculating the bare fermion self-energy in DR to one loop; we present it here for completeness. The Feynman diagram contributing to this two-point Green's function is shown in Fig. \ref{Feynman_Diagram_Zpsi_Zm}.
Denoting by $\Sigma$ the higher-order terms $\mathcal{O} (g^2)$ of the 1-PI amputated Green's function of the fermion field, the inverse full fermion propagator takes the following form: $\langle\psi \bar{\psi}\rangle^{-1} = \imag \slashed{q} + m \openone - \Sigma$. Writing $\Sigma$ in the more useful form: $\Sigma = \imag \slashed{q} \ \Sigma_1 (q^2,m) \ + \ m \ \openone \ \Sigma_2 (q^2,m)$, we present the one-loop results for the functions $\Sigma_1, \Sigma_2$: 
\begin{align} 
\Sigma_1 (q^2,m) = \frac{g^2 C_f}{16 \pi ^2} (\beta -1) \Bigg\{1 + \frac{1}{\varepsilon} + \log \left(\frac{\overline{\mu}^2}{q^2 + m^2}\right) - \frac{m^2}{q^2} \left[1 - \frac{m^2}{q^2} \log \left(1+\frac{q^2}{m^2}\right)\right] \Bigg\} + \mathcal{O} (g^4),
\end{align}
\begin{align} 
\Sigma_2 (q^2,m) = \frac{g^2 C_f}{16 \pi ^2} \Bigg\{2 + (\beta - 4) \Bigg[ 2 + \frac{1}{\varepsilon} + \log \left(\frac{\overline{\mu}^2}{q^2 + m^2}\right) - \frac{m^2}{q^2} \log \left(1+\frac{q^2}{m^2}\right) \Bigg] \Bigg\} + \mathcal{O} (g^4). 
\end{align}

\bigskip

The renormalization conditions for $Z_\psi$ and $Z_m$ in the RI$'$ scheme, using the above notation, take the following perturbative forms:
\be
 Z_\psi^{DR, RI'} = \frac{1}{1 - \Sigma_1} \Big|_{q_\nu = \bar{q}_\nu},
\ee
\be 
Z_m^{DR, RI'} = \frac{1 - \Sigma_1}{1 - \Sigma_2} \Big|_{q_\nu = \bar{q}_\nu}.
\ee

\bigskip

Thus, in the presence of finite fermion masses, the results for the renormalization factors of the fermion field and mass are given below:
\begin{align}
Z_\psi^{DR, RI'} = 1 &+ \frac{g^2 C_f}{16 \pi ^2} (\beta - 1) \Bigg[\frac{1}{\varepsilon} + 1 + \log \left(\frac{\overline{\mu }^2}{\overline{q}^2 + m^2}\right) - \frac{m^2}{\bar{q}^2} \Bigg( 1 - \frac{m^2}{\bar{q}^2} \log \left(1+ \frac{\bar{q}^2}{m^2}\right) \Bigg) \Bigg] \nonumber \\
&+ \mathcal{O} (g^4),
\label{ZpsiRI'}
\end{align}
\begin{align}
Z_m^{DR, RI'} = 1 + \frac{g^2 C_f}{16 \pi ^2} \Bigg[&- \frac{3}{\varepsilon} + \beta - 5 - 3 \log \left(\frac{\overline{\mu }^2}{\overline{q}^2 + m^2}\right) - (\beta - 4) \frac{m^2}{\bar{q}^2} \log \left(1+ \frac{\bar{q}^2}{m^2}\right) \nonumber \\
&+ (\beta - 1) \frac{m^2}{\bar{q}^2} \Bigg( 1 - \frac{m^2}{\bar{q}^2} \log \left(1+ \frac{\bar{q}^2}{m^2}\right) \Bigg) \Bigg] + \mathcal{O} (g^4).
\label{ZmRI'}
\end{align}

We recall that the mass appearing in the above expressions is the renormalized mass, which coincides with the bare mass to this order. The results for $Z_\psi$ and $Z_m$ are in agreement with Ref.\cite{Gracey:2003yr}, in the massless limit and for $\bar{q} = \bar{\mu}$.

\bigskip

The renormalization factors in the $\MSbar$ scheme can be readily inferred from Eqs. \eqref{ZpsiRI'} and \eqref{ZmRI'} by taking only the pole part in epsilon:
\be
Z_\psi^{DR, \MSbar} = 1 + \frac{g^2 C_f}{16 \pi ^2} \frac{1}{\varepsilon} (\beta - 1) + \mathcal{O} (g^4),
\ee
\be
Z_m^{DR, \MSbar} = 1 + \frac{g^2 C_f}{16 \pi ^2} \frac{1}{\varepsilon} (- 3) + \mathcal{O} (g^4).
\ee

\subsubsection{\underline{\textbf{Renormalization factors of Wilson-line operators}}}
Now, we have all the ingredients for the extraction of renormalization factors of Wilson-line operators in the RI$'$ and $\MSbar$ schemes. By writing $Z_{\psi_f}$ and $\Lambda_\Gamma$ in the form:
\be 
Z_{\psi_f}^{DR,Y} = 1 + g^2 z_{\psi_f}^Y + \mathcal{O} (g^4),
\ee
\be 
\Lambda_{\Gamma_i} = \Lambda_{\Gamma_i}^{\text{tree}} + \Lambda_{\Gamma_i}^{\text{1-loop}} + \mathcal{O} (g^4), \qquad (i = 1,2),
\ee
where\footnote{The Green's functions $\Lambda_{\Gamma_i}^{\text{1-loop}}$ also contain additional Dirac structures [see Eqs. \eqref{LambdaSd1} - \eqref{Lambdad4}], which do not contribute to the evaluation of renormalization factors $Z_\Gamma$ in the $\MSbar$ scheme, as they are $\mathcal{O} (\varepsilon^0)$ terms, nor in RI$'$, as the trace in Eq. \eqref{RC_Lambda_RIprime2} gives zero.} 
\be 
\Lambda_{\Gamma_i}^{\text{1-loop}} = g^2 \sum_{j=1}^2 \lambda_{ij} \ \Lambda^{\text{tree}}_{\Gamma_j} + \cdots, 
\qquad
\lambda_{ij} = {1\over 4N_c}\,{1\over g^2}\,  {\rm Tr}\Big[\Lambda_{\Gamma_i}^{\text{1-loop}} \, (\Lambda_{\Gamma_j}^{\rm tree})^\dagger\Big],
\ee
the condition for the renormalization of Wilson-line operators in the RI$'$ scheme, up to one loop, reads 
\be 
\Big[Z^{DR, RI'}_{\Gamma_1, \Gamma_2}\Big]_{ij} = \delta_{ij} + g^2 \delta_{ij} \Big(\frac{1}{2} z_{\psi_f}^{RI'} + \frac{1}{2} z_{\psi_{f'}}^{RI'} + \lambda_{ii}\Big\vert_{q_\nu = \bar{q}_\nu}\Big) + g^2 (1 - \delta_{ij}) \lambda_{ij}\Big\vert_{q_\nu = \bar{q}_\nu}.
\label{Z_Gamma_RI'}
\ee 
The equivalent expression for $Z_\Gamma^{DR,\MSbar}$ follows from Eq. \eqref{Z_Gamma_RI'}, by keeping in $\lambda_{ij}$ only pole parts in epsilon; the latter appear only for $i=j$, leading to
\be
 Z_{\Gamma_i}^{DR, \MSbar} = 1 + g^2 \Big(\frac{1}{2} z_{\psi_f}^{\MSbar} + \frac{1}{2} z_{\psi_{f'}}^{\MSbar} + \lambda_{ii} \Big\vert_{1/\varepsilon} \Big). 
\ee

\bigskip

Our final results are presented below. In the $\MSbar$ scheme, the renormalization factors of operators have the form:
\be 
Z^{DR, \MSbar}_\Gamma = 1 + \frac{g^2\, C_f}{16\,\pi^2} \, \frac{3}{\varepsilon} + \mathcal{O} (g^4),
\ee
in agreement with Refs.\cite{Stefanis:1983ke, Dorn:1986dt, Chetyrkin:2003vi}. As we observe, they are independent of operator $\Gamma$, fermion masses, Wilson-line length $z$, and gauge parameter $\beta$. In RI$'$, the renormalization factors  are given with respect to the conversion factors, which are presented in the next section: 
\be 
\Big[Z^{DR, RI'}_{\Gamma_1, \Gamma_2}\Big]_{ij} =  \Big[\mathcal{C}_{\Gamma_1, \Gamma_2}^{\MSbar,{\rm RI}'}\Big]_{ij} + \frac{g^2\, C_f}{16\,\pi^2} \, \frac{3}{\varepsilon} \delta_{ij} + \mathcal{O} (g^4).
\ee 
The above relation stems from the one-loop expression of Eq. \eqref{ConvFactors}.

\subsection{Conversion factors}
We present below our results for all the matrix elements of $2 \times 2$ conversion factors in a compact way. We use the same notation as in Sec. III A for bare Green's functions; the only difference is that the Feynman parameter integrals, appearing here, depend on the RI$'$ scale $\bar{q}$ instead of the external momentum $q$: 

\be
\begin{split}
\left[C_{S,V_{\mu}} \right]_{11} = \ 1+\frac{g^2 C_f}{16 \pi ^2} \Bigg\{&7 -3 \beta + 2 (\beta + 2) \gamma_E +2 (\beta - 1)\Big(f_{11}+f_{12}\Big) - (\beta - 4) \Big( 4 h_1 - \left| z\right| h_4 \Big) \\
&+ 3 \log\left(\frac{\overline{\mu }^2}{\overline{q}^2}\right) + (\beta +2) \log \left(\frac{1}{4} z^2 \overline{q}^2\right) +\frac{1}{2} (\beta - 1) \Bigg[ - \frac{m_1^2}{\bar{q}^2}-\frac{m_2^2}{\bar{q}^2} \\
&+\frac{m_1^2}{\bar{q}^2} \left(2 +\frac{m_1^2}{\bar{q}^2}\right) \log \left(1+ \frac{\bar{q}^2}{m_1^2}\right) +\frac{m_2^2}{\bar{q}^2} \left(2 +\frac{m_2^2}{\bar{q}^2} \right) \log \left(1+\frac{\bar{q}^2}{m_2^2}\right) \\
&+ \log \left(1 + \frac{m_1^2}{\bar{q}^2}\right)+\log \left(1 + \frac{m_2^2}{\bar{q}^2}\right) \Bigg] + 2 \left| z\right| m_1 m_2 (\beta - 2) h_5 \\
&+ \beta \left| z\right| \Big(\bar{q}^2 - m_1 m_2 \Big) \Big(2 h_6 - \bar{q}^2 h_8\Big) - \bar{q}^2 \left| z\right| \Big(\beta h_5 + (\beta - 4) h_7 \Big) \\ 
&+\beta \bar{q}^2 \Big[ (m_1^2 + \bar{q}^2) \ g_{41}+ (m_2^2 + \bar{q}^2) \ g_{42}- (\bar{q}^2 - \bar{q}_{\mu}^2) \Big(g_{51}+g_{52}\Big) \Big] \\
&+\frac{1}{2} \beta z^2 \Big(\bar{q}^2 -  m_1 m_2 \Big) \Big(h_2 - \bar{q}^2 h_3\Big)  \\
&-2 \ \imag \ \bar{q}_{\mu} \Big(g_{11}+g_{12}\Big) +\imag z \bar{q}_{\mu} \Big[\beta \Big(f_{21}+  f_{22}\Big) - 2 (\beta - 4) h_2\Big] \\
&+ \imag \beta \bar{q}^2 \bar{q}_{\mu} \Big[ g_{31}+g_{32} - z \Big(f_{31}+f_{32}\Big) \Big] \\
&+ \imag \beta  z \left| z\right| \bar{q}_{\mu} \Big(\bar{q}^2-m_1 m_2\Big) \Big(h_5-h_6-h_7\Big) \Bigg\} + \mathcal{O} (g^4), 
\end{split} 
\ee

\be
\begin{split}
\left[C_{S,V_{\mu}} \right]_{12} = \ \frac{g^2 C_f}{16 \pi ^2} \Bigg\{&-\beta z \Big( m_1 f_{21} + m_2 f_{22} \Big) - \beta \bar{q}_{\mu}^2 \Big[m_1 \Big(g_{31} - z f_{31}\Big)+ m_2 \Big(g_{32} - z f_{32}\Big) \Big] \\
&+\imag \beta \bar{q}_{\mu} \Big[ m_1 (m_1^2 + \bar{q}^2) g_{41} + m_2 (m_2^2 + \bar{q}^2) g_{42} - (\bar{q}^2 - \bar{q}_{\mu}^2) \Big(m_1 g_{51} + m_2 g_{52}\Big)\Big] \\
&+ 2 \Big(m_1 g_{11}+m_2 g_{12}\Big) + \Big(m_1+m_2\Big) \Bigg[ (\beta +2) z h_1  -\imag \beta  \bar{q}^2 \left| z\right| \bar{q}_{\mu} h_8 \\
&+ \imag \left| z\right| \bar{q}_{\mu} \Big((\beta - 2) h_5 + 2 h_6\Big) +\frac{1}{2} \imag \beta  z^2 \bar{q}_{\mu} \Big(h_2-\bar{q}^2 h_3\Big) \\ 
&-\beta  \bar{q}^2 z \left| z\right| \Big(h_5-h_6-h_7\Big) \Bigg] \Bigg\} + \mathcal{O} (g^4), \\
\end{split} 
\ee

\be
\begin{split}
\left[C_{S,V_{\mu}} \right]_{21} = \left[C_{S,V_{\mu}} \right]_{12} + \frac{g^2 C_f}{16 \pi ^2}\Big(m_1+m_2\Big) \Bigg\{& -6 z h_1 -3 \imag (\beta - 2) \left| z\right| \bar{q}_{\mu} \Big(h_5-h_6\Big) \\
& +\beta  z \left| z\right|  \left(\bar{q}^2-\bar{q}_{\mu}^2\right)\Big(h_5-h_6-h_7\Big)\Bigg\} + \mathcal{O} (g^4),
\end{split} 
\ee

\be
\begin{split}
\left[C_{S,V_{\mu}} \right]_{22} = \left[C_{S,V_{\mu}} \right]_{11} + \frac{g^2 C_f}{16 \pi ^2} \Bigg\{&-12 h_1 -12 \imag z \bar{q}_{\mu} h_2 + 3 \left| z\right|  \Big[2 h_4 - m_1 m_2 \Big((\beta - 2) h_5-\beta h_6\Big)\Big] \\
&-2 \beta \left| z\right| m_1 m_2 (\bar{q}^2 - \bar{q}_{\mu}^2) h_8 +\left| z\right|  (\bar{q}^2 + 2\bar{q}_{\mu}^2) \Big[ \beta \Big(h_5-h_6\Big)-2 h_7\Big] \\
&-\beta  z^2 \left(\bar{q}^2 - \bar{q}_{\mu}^2\right)\Big(h_2+m_1 m_2 h_3\Big) \Bigg\} + \mathcal{O} (g^4),
\end{split} 
\ee

\begin{gather} 
[C_{P,A_{\mu}}]_{ij} = [C_{S,V_{\mu}}]_{ij} \{h_k \mapsto (-1)^{1 + \delta_{ij}} h_k, m_1 \mapsto - m_1\} \\
(\textrm{where} \ i,j = 1,2 \ \textrm{and} \ k=1,2,\cdots ,8), \nonumber
\end{gather}

\be
\begin{split}
\left[C_{V_{\nu},T_{\mu \nu}} \right]_{11} = \left[C_{P,A_{\mu}} \right]_{11} + \frac{g^2 C_f}{16 \pi ^2} \Bigg\{&-12 h_1 -4 \imag z \bar{q}_{\mu} h_2 + \left| z\right| (\bar{q}^2 + 2 \bar{q}_{\nu}^2) \Big[\beta \Big(h_5-h_6\Big)-2 h_7\Big] \\
&+\left| z\right| \Big[2 h_4+m_1 m_2 \Big((\beta - 2) h_5-\beta  h_6 + 2 \beta \bar{q}_{\nu}^2 h_8 \Big)\Big] \\
&-\beta  z^2 \bar{q}_{\nu}^2 \Big(h_2 - m_1 m_2 h_3\Big) \Bigg\} + \mathcal{O} (g^4),
\end{split}
\ee

\be
\begin{split}
\left[C_{V_{\nu},T_{\mu \nu}} \right]_{12} = -\left[C_{P,A_{\mu}} \right]_{12} + \frac{g^2 C_f}{16 \pi ^2} \Big(m_1-m_2\Big)\Bigg\{&2 z h_1 + \imag (\beta - 2) \left| z\right|  \bar{q}_{\mu}\Big(h_5-h_6\Big) \\
&-\beta  z \left| z\right| \bar{q}_{\nu}^2 \Big(h_5-h_6-h_7\Big) \Bigg\} + \mathcal{O} (g^4),
\end{split}
\ee

\be
\begin{split}
\left[C_{V_{\nu},T_{\mu \nu}} \right]_{21} = -\left[C_{P,A_{\mu}} \right]_{21} - \frac{g^2 C_f}{16 \pi ^2} \Big(m_1-m_2\Big)\Bigg\{&2 z h_1 + \imag (\beta - 2) \left| z\right|  \bar{q}_{\mu}\Big(h_5-h_6\Big) \\
&-\beta  z \left| z\right| \bar{q}_{\nu}^2 \Big(h_5-h_6-h_7\Big) \Bigg\} + \mathcal{O} (g^4),
\end{split}
\ee

\be
\begin{split}
\left[C_{V_{\nu},T_{\mu \nu}} \right]_{22} = \left[C_{P,A_{\mu}} \right]_{22} + \frac{g^2 C_f}{16 \pi ^2} \Bigg\{&-4 h_1 +4 \imag z \bar{q}_{\mu} h_2 + \left| z\right| (\bar{q}^2 - 2 \bar{q}_{\nu}^2) \Big[\beta \Big(h_5-h_6\Big)-2 h_7\Big] \\
&-\left| z\right| \Big[2 h_4+m_1 m_2 \Big((\beta - 2) h_5-\beta  h_6 + 2 \beta \bar{q}_{\nu}^2 h_8 \Big)\Big] \\
&+\beta  z^2 \bar{q}_{\nu}^2 \Big(h_2 - m_1 m_2 h_3\Big) \Bigg\} + \mathcal{O} (g^4),
\end{split}
\ee

\begin{gather} 
[C_{A_{\nu},T_{\rho \sigma}}]_{ij} = {(- \varepsilon_{\mu \nu \rho \sigma})}^{1 + \delta_{ij}} \ [C_{V_{\nu},T_{\mu \nu}}]_{ij} \{h_k \mapsto (-1)^{1 + \delta_{ij}} h_k, m_1 \mapsto - m_1\} \\
(\textrm{where} \ i,j = 1,2 \ \textrm{and} \ k=1,2,\cdots ,8; \ \varepsilon_{\mu \nu \rho \sigma} \  \textrm{is the Levi-Civita tensor}, \ \varepsilon_{1234} = 1). \nonumber  
\end{gather}

Our results are in agreement with Ref.\cite{Constantinou:2017sej} in the massless limit\footnote{Checking agreement is quite nontrivial; it requires the elimination of certain integrals over Feynman parameters, integration by parts, as well as the interchange of the limit operation with integration.}. A consequence of the above relations is that, in the case of equal quark masses $m_1 = m_2$, the nondiagonal matrix elements of $C_{P, A_\mu}$ and $C_{V_\nu,T_{\mu \nu}}$ vanish. Also, the matrix elements of conversion factors satisfy the following reflection relation with respect to $z$: 
\be
{[C_{\Gamma_1, \Gamma_2} (\bar{q}, z, m_1, m_2)]}_{ij} = {(-1)}^{1 + \delta_{ij}} {[C_{\Gamma_1, \Gamma_2}^\ast (\bar{q}, -z, m_1, m_2)]}_{ij}. 
\label{refrel} 
\ee
This means that the real part of diagonal (nondiagonal) matrix elements is an even (odd) function of z, while the imaginary part is odd (even).   

\section{Graphs}

In this section, we illustrate our results for conversion factors by selecting certain values of the free parameters used in simulations. To this end, we plot the real and imaginary parts of the conversion factor matrix elements as a function of Wilson-line length, z. For input, we employ certain parameter values, used by ETMC in the ensemble of dynamical $N_f = 2+1+1$ twisted mass fermions of Ref.\cite{Alexandrou:2016jqi}; i.e., we set\footnote{A most natural choice for the coupling constant would be its $\MSbar$ value, even though the choice of bare vs renormalized coupling constant should, in principle, be irrelevant for one-loop results, such as the ones we plot in this section. Nevertheless, these plots are meant to reveal some salient features of the conversion factors, which certainly are not affected by selecting $g^2 \sim 3.77$ ($\MSbar$) rather than $g^2 = 3.077$ (lattice); indeed, given the simple linear dependence on $g^2$ of the
quantities plotted, the effect of a change in $g^2$ can be inferred by inspection. For precise quantitative values of the conversion factors, one should of course refer to our results in algebraic form, presented in Sec. III.} $g^2 = 3.077$, $ \beta = 1$ (Landau gauge), $N_c = 3$, $\bar{\mu} = 2$ GeV, and $\bar{q} = \frac{2 \pi}{32 a} (n_z,0,0,\frac{n_t}{2}+\frac{1}{4})$, for $a = 0.082$ fm (lattice spacing), $n_z = 4$, and $n_t = 8$ (the Wilson line is taken to lie in the z direction, which, by convention, is denoted by $\mu = 1$). Expressed in GeV, $\bar{q} = (1.887,0,0,2.048)$ GeV. To examine the impact of finite quark masses on the conversion factors, we plot six different cases of external quark masses: \\ 
\phantom{a}\hspace{0.5cm} 1. massless quarks ($m_1 = m_2 = 0$) \\
\phantom{a}\hspace{0.5cm} 2. $m_1 = m_2 = 13.2134$ MeV, corresponding to the bare twisted mass used in Ref.\cite{Alexandrou:2016jqi} \\
\phantom{a}\hspace{0.5cm} 3. one up and one strange quark ($m_1 = 2.3$ MeV, $m_2 = 95$ MeV) \\
\phantom{a}\hspace{0.5cm} 4. two strange quarks ($m_1 = m_2 = 95$ MeV) \\
\phantom{a}\hspace{0.5cm} 5. one up and one charm quark ($m_1 = 2.3$ MeV, $m_2 = 1275$ MeV) \\ 
\phantom{a}\hspace{0.5cm} 6. two charm quarks ($m_1 = m_2 = 1275$ MeV). \\
As regards the $\bar{q}$ dependence, we have not included further graphs for the sake of conciseness; however, using a variety of values for the components of $\bar{q}$, we find no significant difference. More quantitative assessments can be directly obtained from our algebraic results.

\bigskip

\begin{figure}[!thb] 
  \centering
  \vspace{-0.67cm}
  \includegraphics[width=7.9cm,clip]{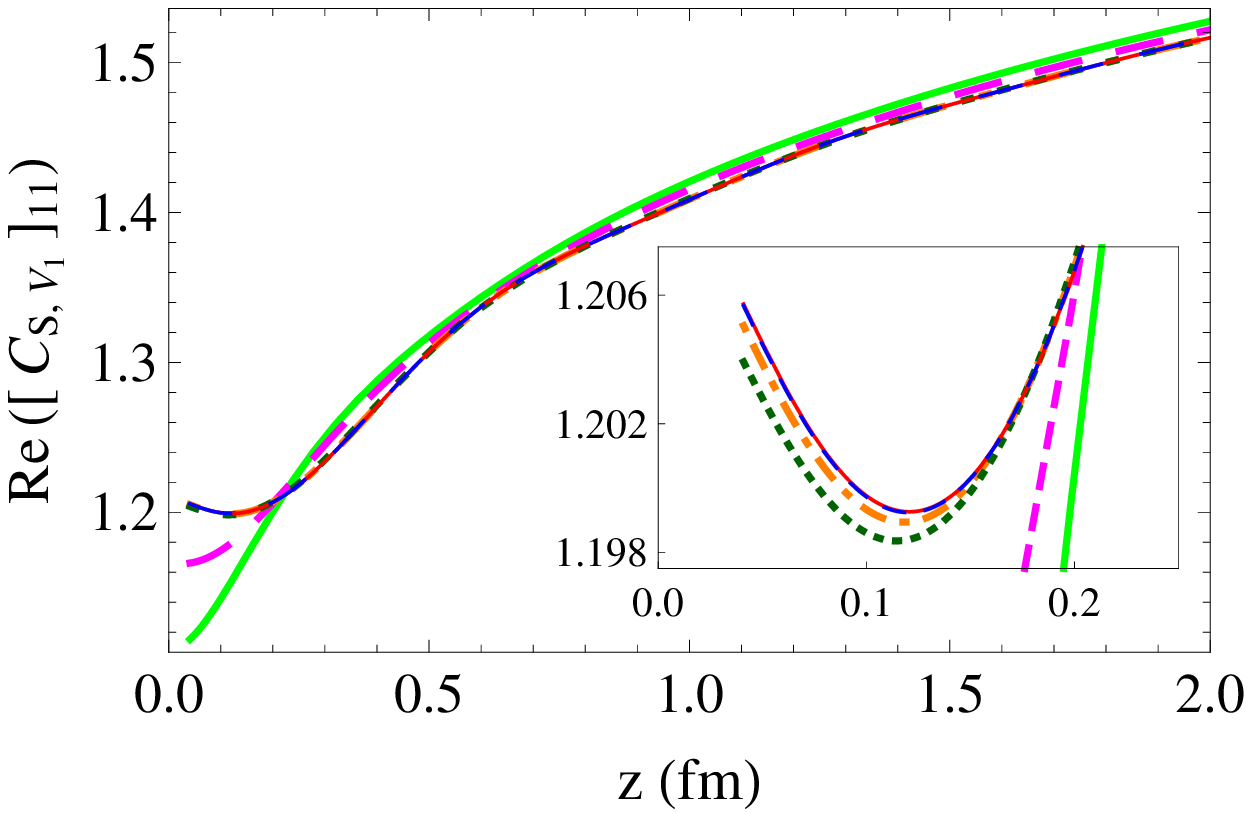} \hspace{0.2mm}
  \includegraphics[width=8.3cm,clip]{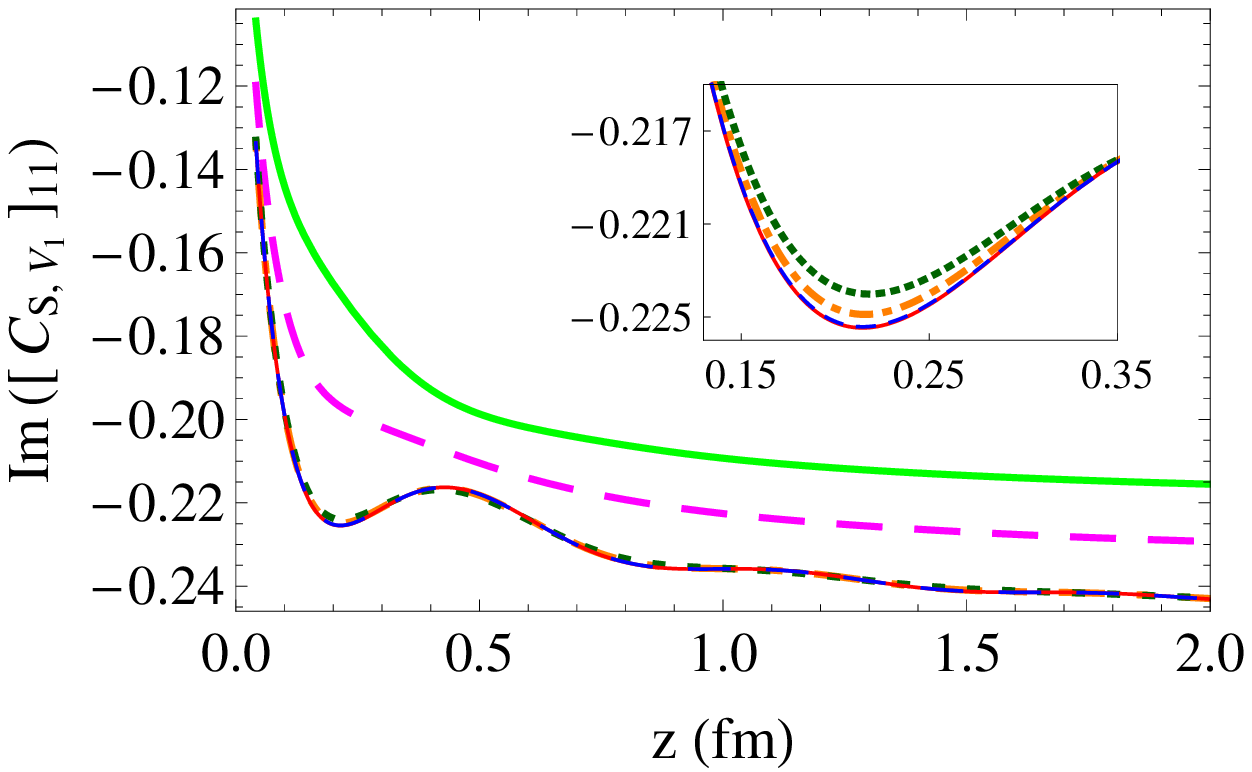} \\
  \vspace{0.27cm}
  \includegraphics[width=7.9cm,clip]{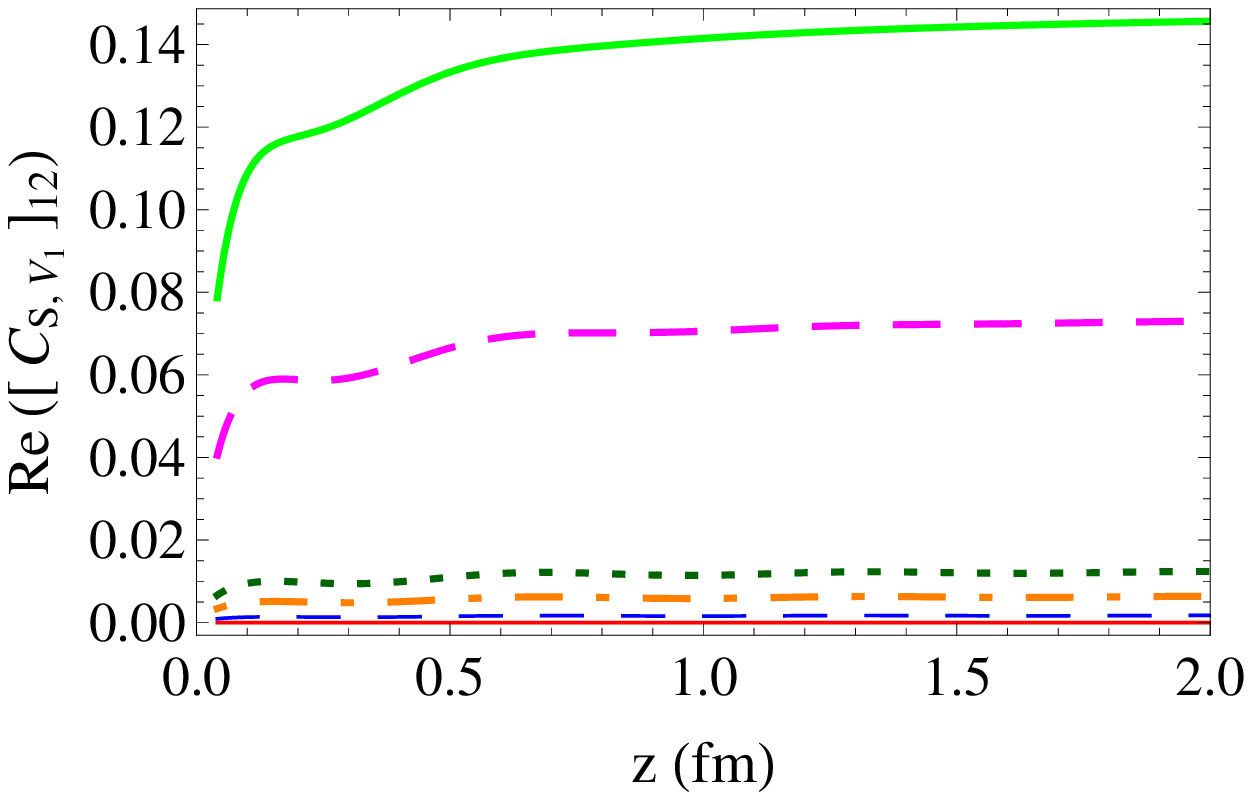} \hspace{0.2mm}
  \includegraphics[width=8.3cm,clip]{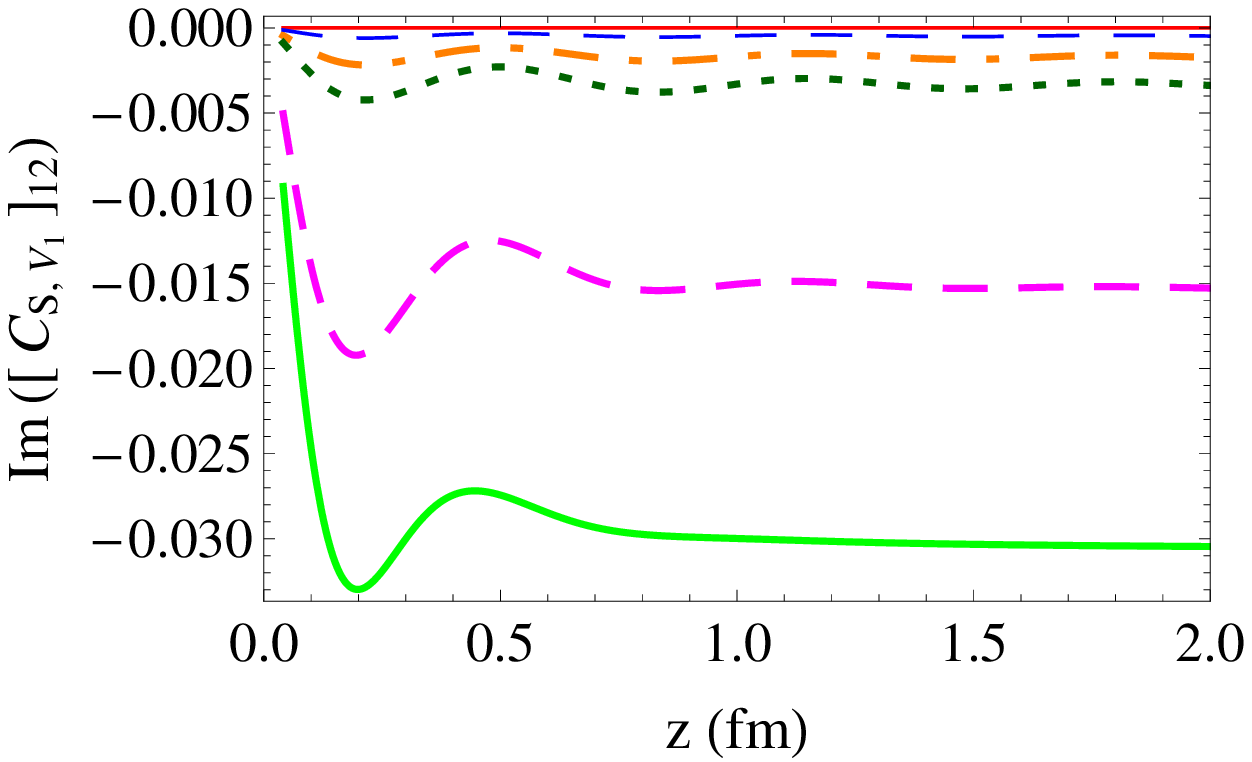} \\ 
  \vspace{0.27cm}
  \includegraphics[width=7.9cm,clip]{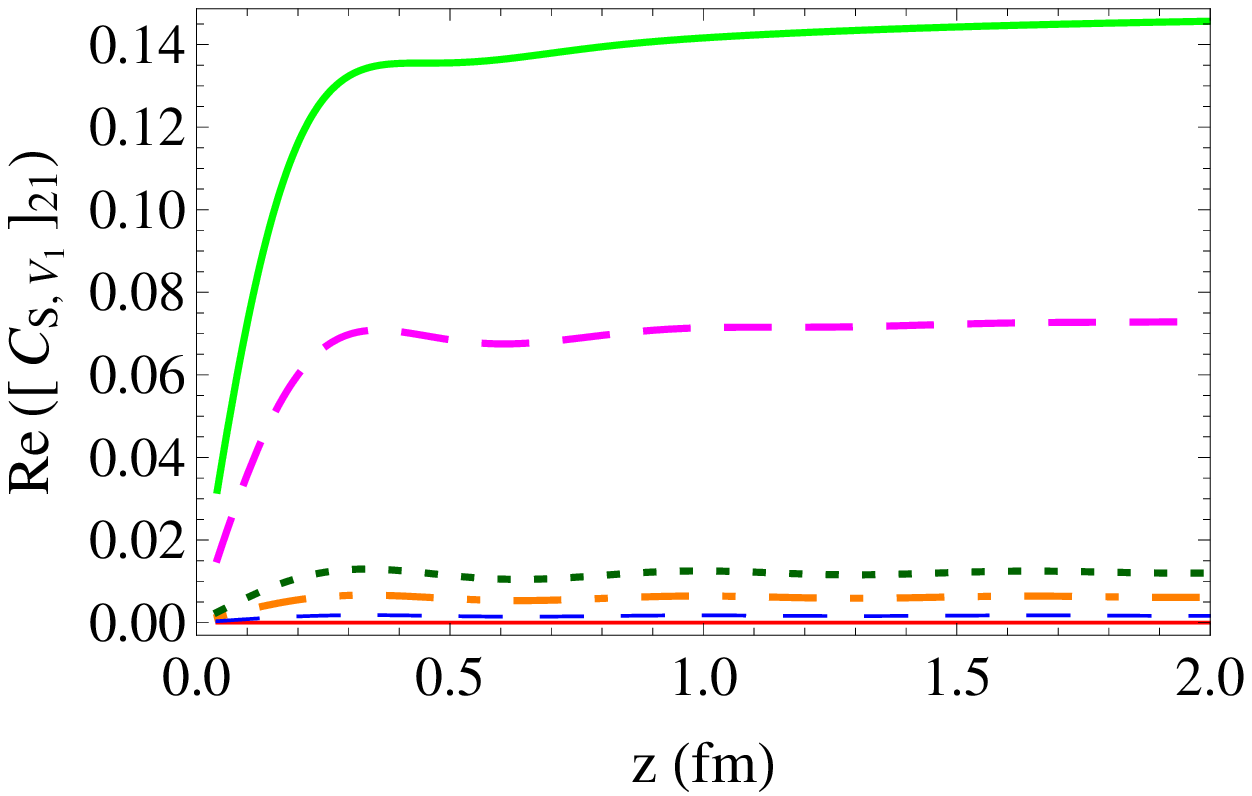} \hspace{0.2mm}
  \includegraphics[width=8.3cm,clip]{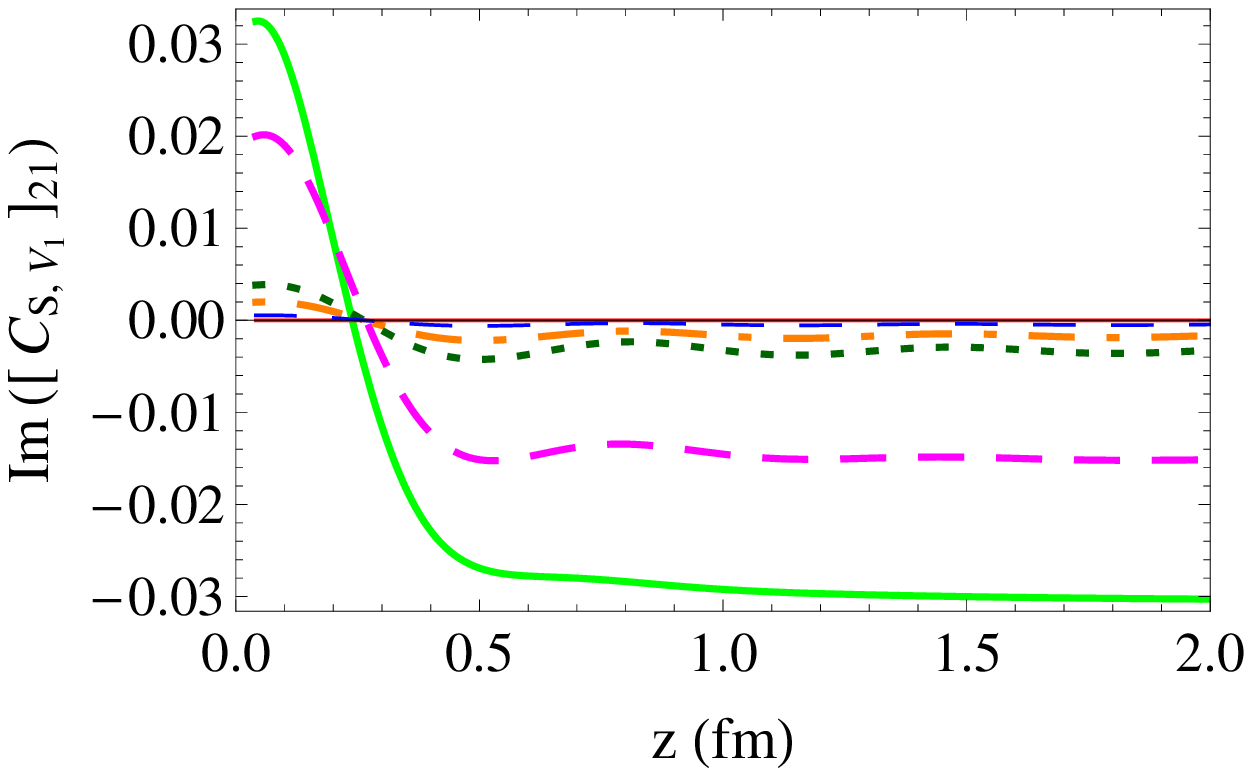} \\
  \vspace{0.27cm}
  \includegraphics[width=7.9cm,clip]{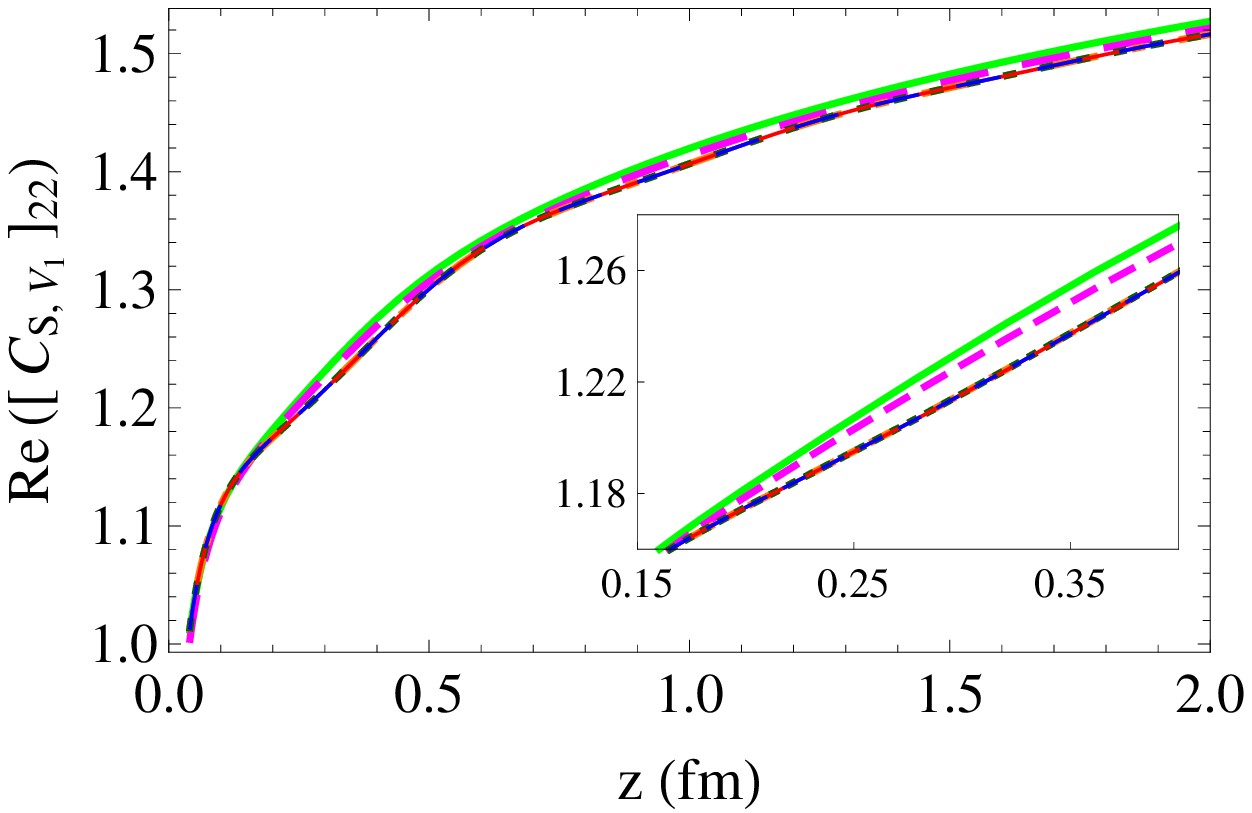} \hspace{0.2mm}
  \includegraphics[width=8.3cm,clip]{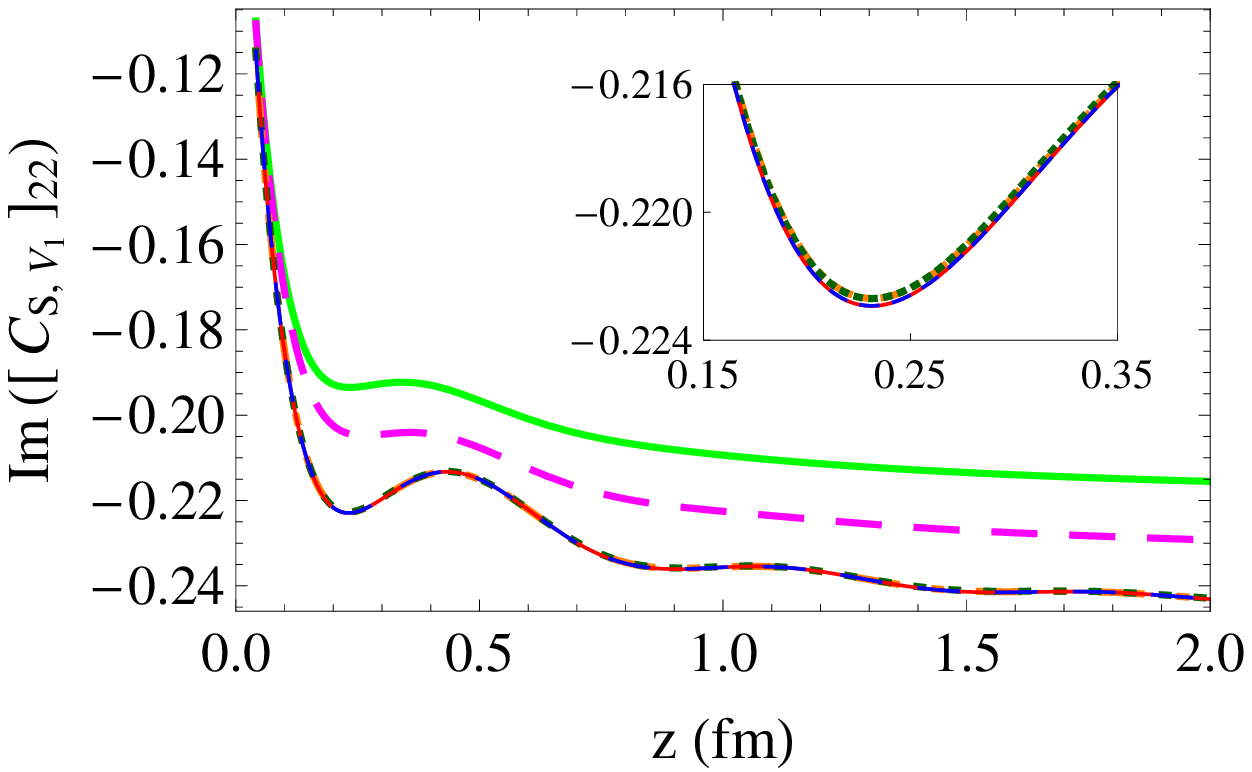} \\
  \vspace{0.15cm}
  \includegraphics[width=14cm,clip]{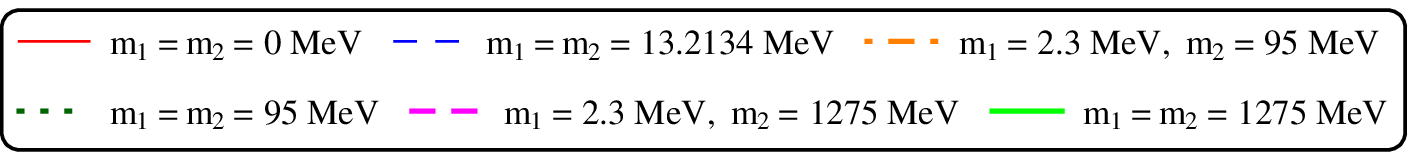}
  \vspace{-0.23cm}
  \caption{Real (left panels) and imaginary (right panels) parts of the conversion factor matrix elements for the operator pair ($S$, $V_1$) as a function of z, for different values of quark masses \qquad \qquad [$g^2 = 3.077$, $\quad \beta = 1$, $\quad N_c = 3$, $\quad \bar{\mu} = 2$ GeV, $\quad \bar{q} = \frac{2 \pi}{32 \ (0.082 \ {\scriptsize \textrm{fm}})} (4,0,0,\frac{17}{4})$].}
  \label{plot1}
  \end{figure}

\begin{figure}[!thb] 
  \centering
  \vspace{-0.67cm}
  \includegraphics[width=7.9cm,clip]{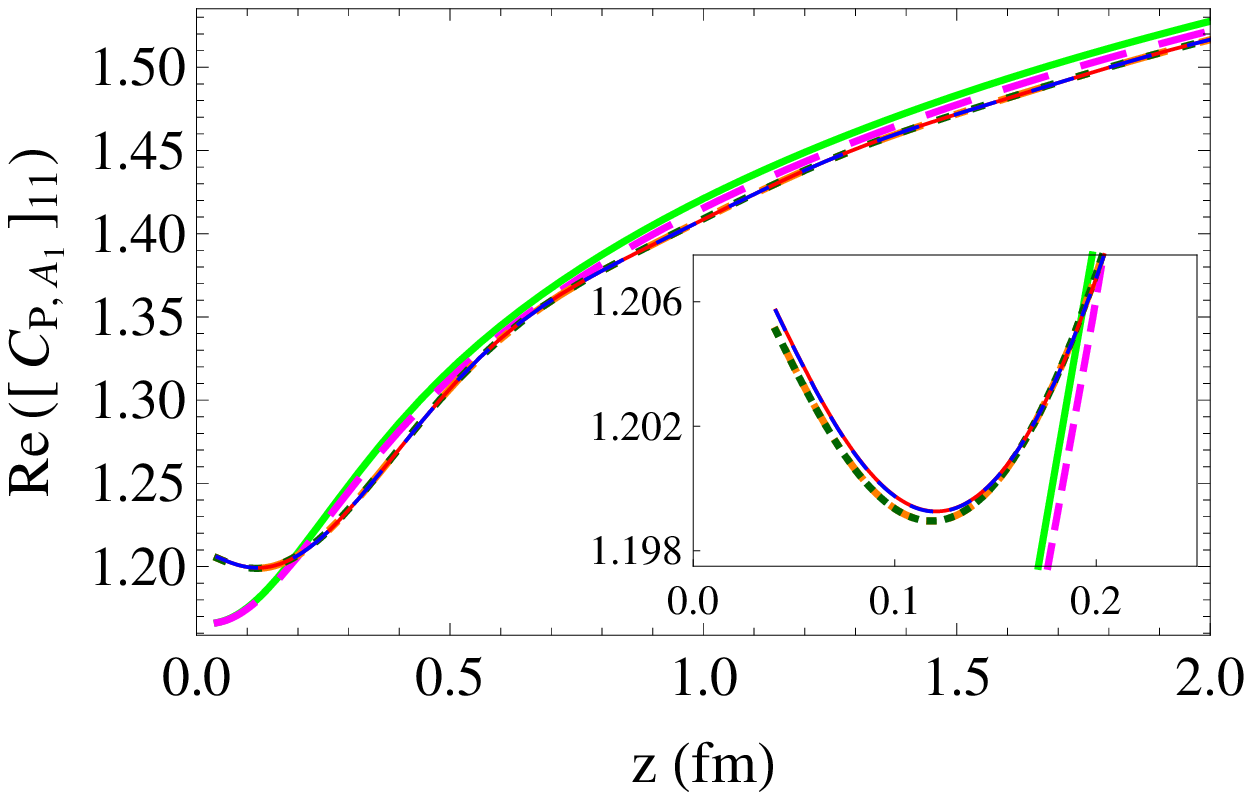} \hspace{0.2mm}
  \includegraphics[width=8.3cm,clip]{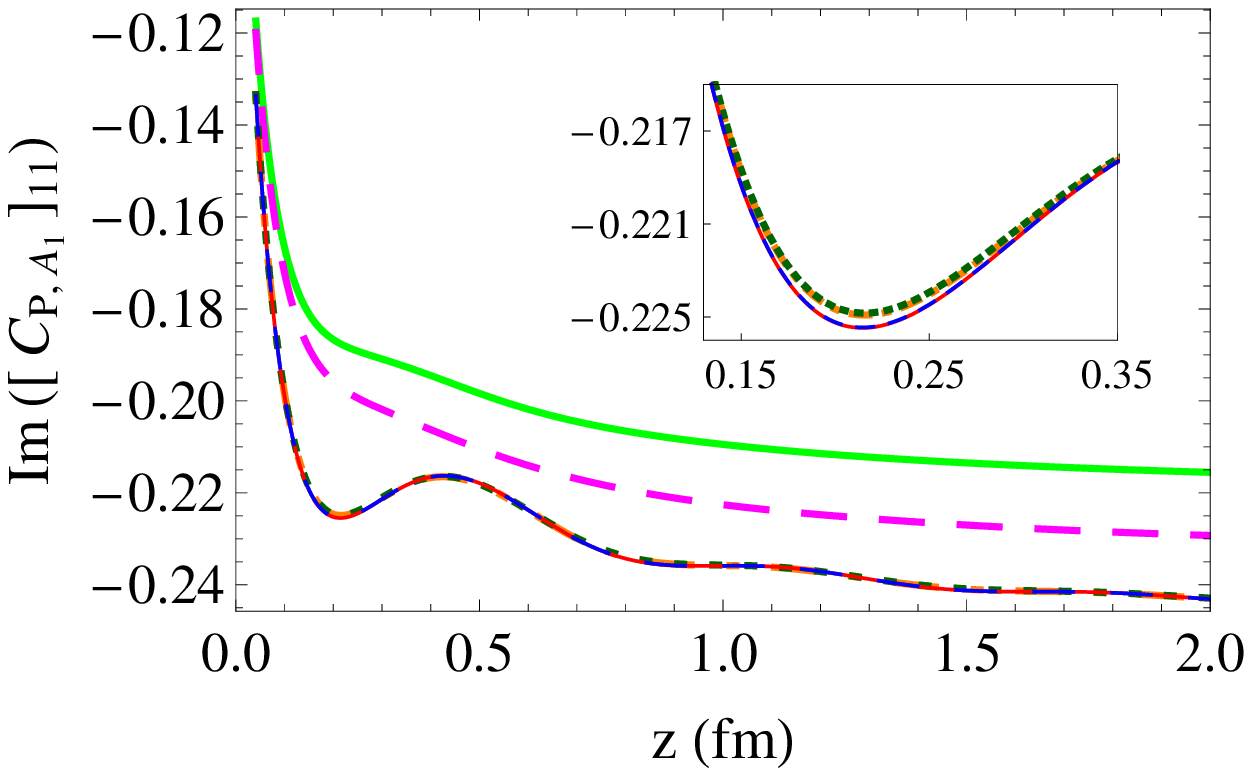} \\
  \vspace{0.27cm}
  \includegraphics[width=7.9cm,clip]{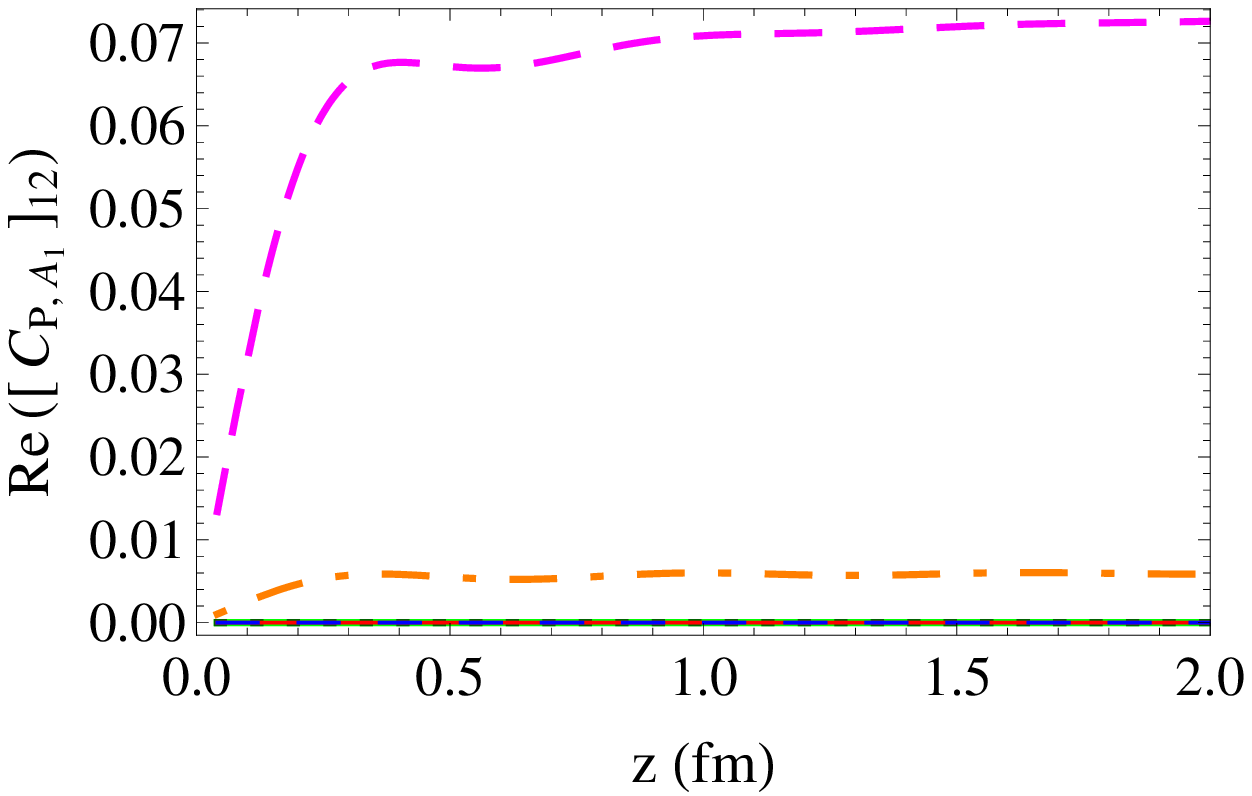} \hspace{0.2mm}
  \includegraphics[width=8.3cm,clip]{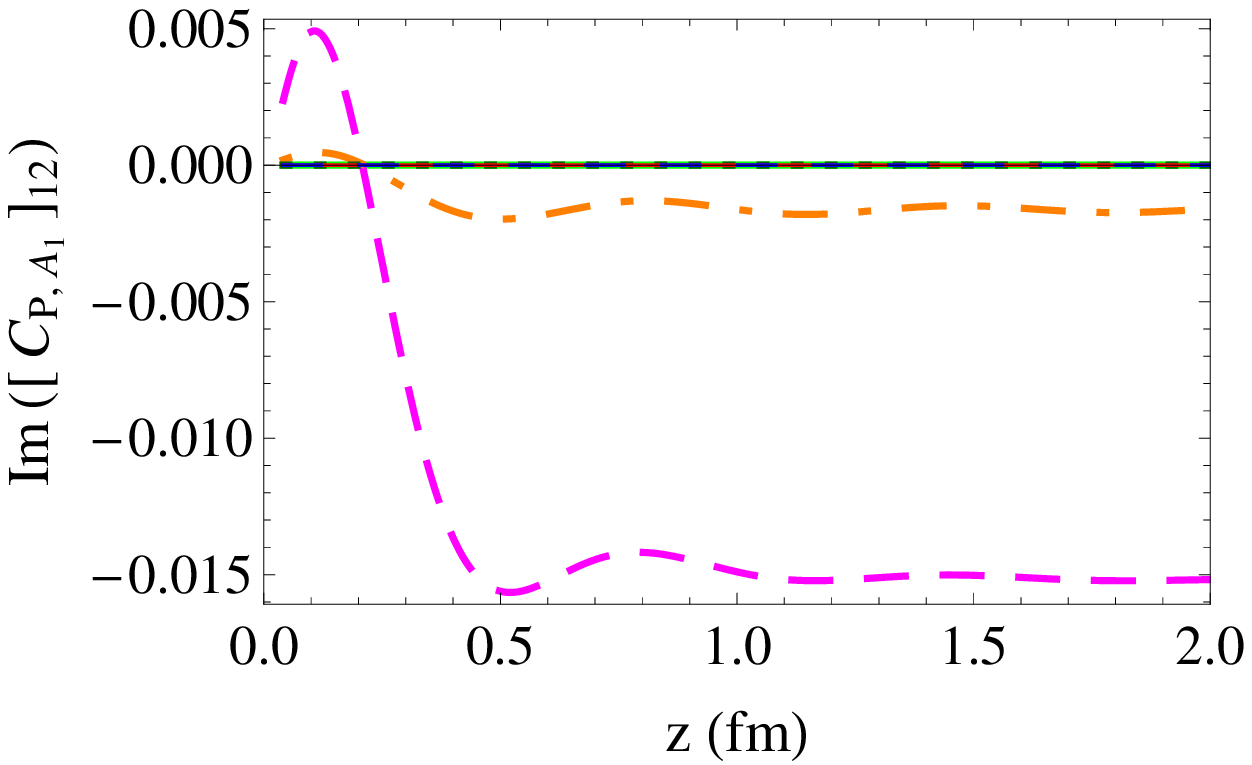} \\ 
  \vspace{0.27cm}
  \includegraphics[width=7.9cm,clip]{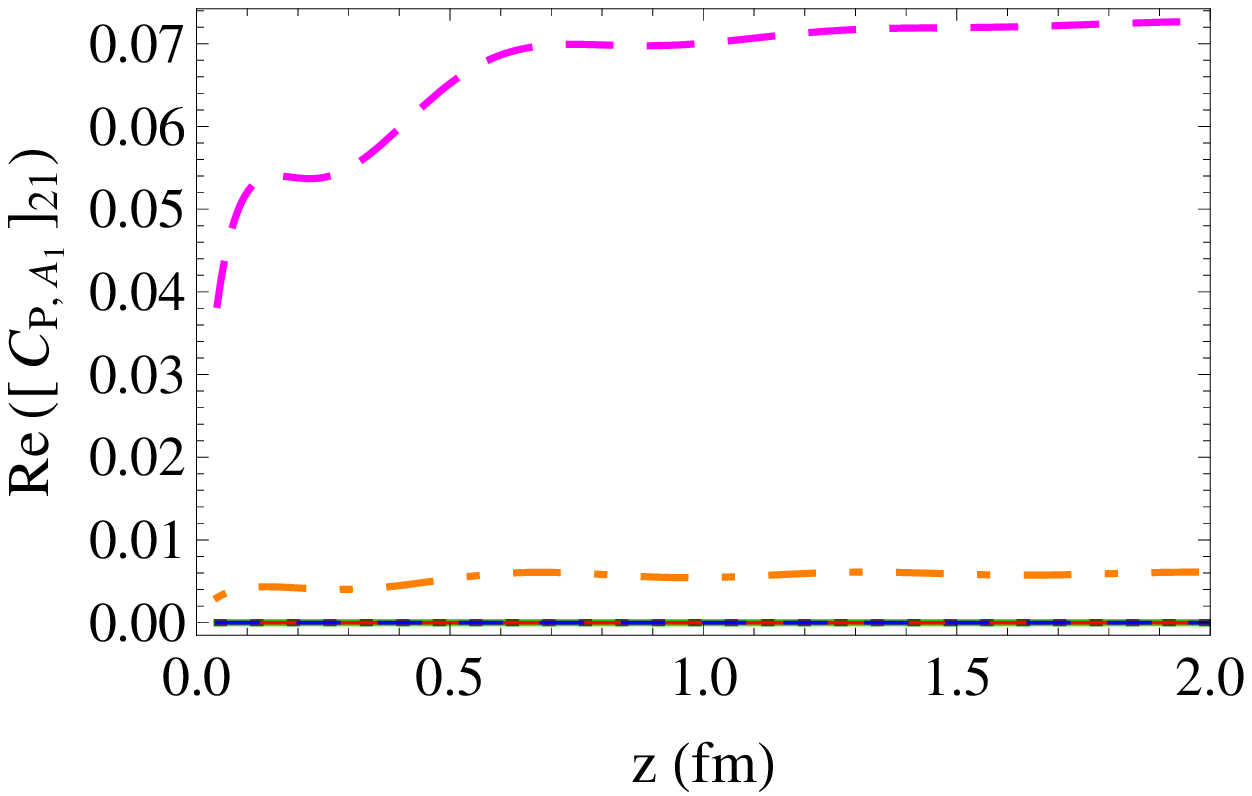} \hspace{0.2mm}
  \includegraphics[width=8.3cm,clip]{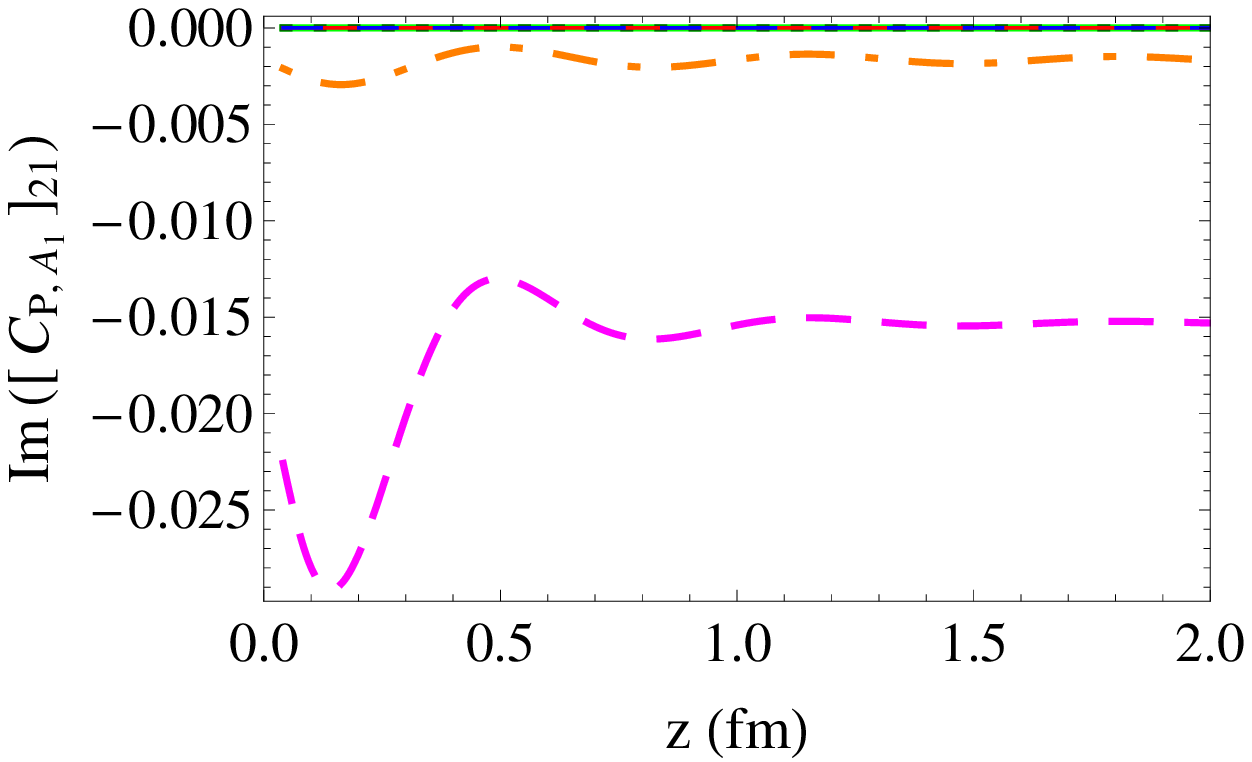} \\
  \vspace{0.27cm}
  \includegraphics[width=7.9cm,clip]{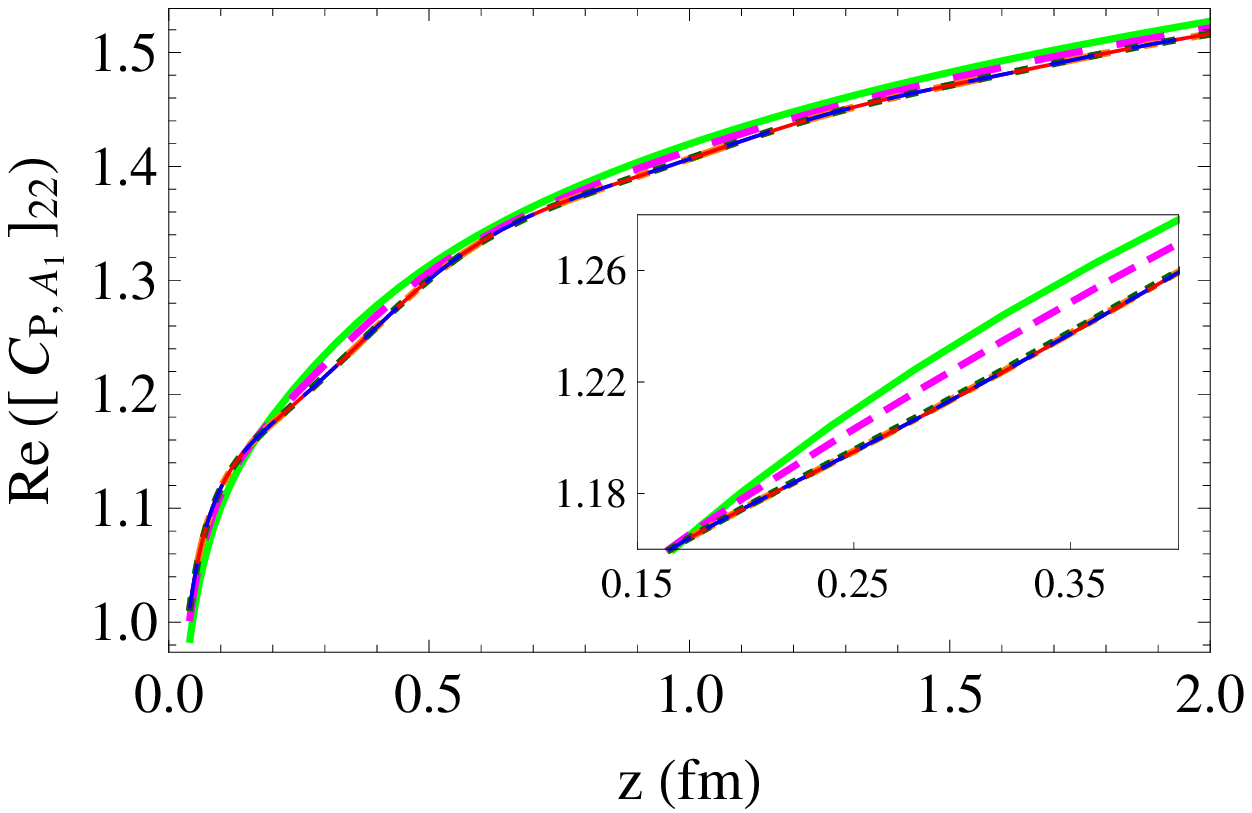} \hspace{0.2mm}
  \includegraphics[width=8.3cm,clip]{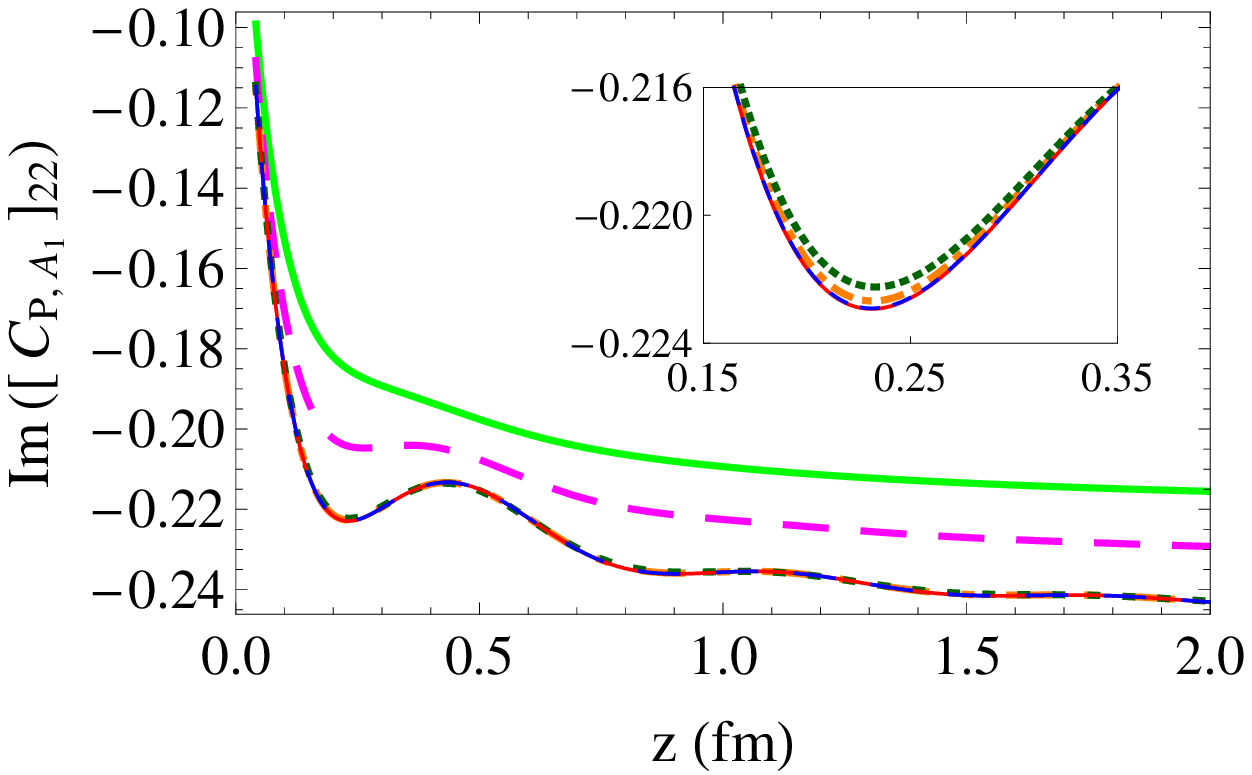} \\
  \vspace{0.15cm}
  \includegraphics[width=14cm,clip]{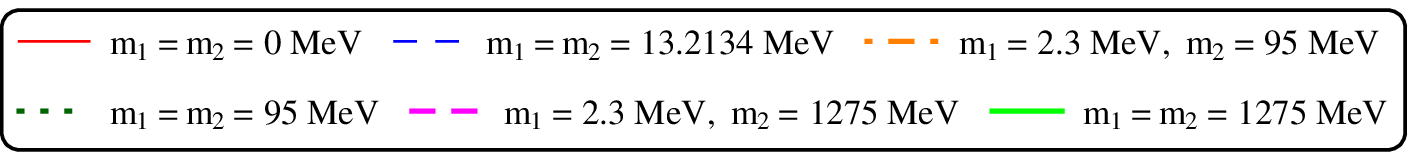}
  \vspace{-0.23cm}
  \caption{Real (left panels) and imaginary (right panels) parts of the conversion factor matrix elements for the operator pair ($P$, $A_1$) as a function of z, for different values of quark masses [$g^2 = 3.077$, $\quad \beta = 1$, $\quad N_c = 3$, $\quad \bar{\mu} = 2$ GeV, $\quad \bar{q} = \frac{2 \pi}{32 \ (0.082 \ {\scriptsize \textrm{fm}})} (4,0,0,\frac{17}{4})$].}
  \label{plot2}
  \end{figure}

In Figs. \ref{plot1} and \ref{plot2}, we present graphs of some representative conversion factors ($C_{S,V_1}$, $C_{P,A_1}$) for the six cases of external quark masses. The plots are given only for positive values of $z$, since the behavior of conversion factors for negative values follows the reflection relation of Eq.\eqref{refrel}. We observe that the real part of the conversion factor matrix elements is an order of magnitude larger than the imaginary part and that the diagonal elements are an order of magnitude larger than the nondiagonal elements. Also, for increasing values of $z$, the real part of diagonal elements tends to increase, while the imaginary part as well as the real part of nondiagonal elements tend to stabilize. Diagonal elements are almost equal to each other, as regards both their real and imaginary parts; a similar behaviour is also exhibited by the nondiagonal elements. Further, the diagonal elements of $C_{S,V_1}$ and $C_{P,A_1}$ behave almost identically, while the nondiagonal elements have different behavior; this is to be expected, given that the cases of equal masses give zero nondiagonal elements for $C_{P,A_1}$. Comparing the six cases, we deduce that the impact of mass becomes significant when we include a strange or a charm quark; the presence of a strange quark causes changes of order $0.005 - 0.01$ for real parts, and $0.001 - 0.003$ for imaginary parts, while the presence of a charm quark causes changes of order $0.07 - 0.14$ for real parts and $0.015 - 0.03$ for imaginary parts. On the contrary, the cases of massless quarks and $m_1 = m_2 = 13.2134$ MeV are almost coincident. Therefore, we conclude that, for quark masses quite smaller than the strange quark mass, we may ignore the mass terms in our calculations, while for larger values, the mass terms are significant.

\bigskip

Regarding the convergence of the perturbative series, we note that one-loop contributions are a small fraction of the tree-level values, which is a desirable indication of stability. Nevertheless, given that these contributions are not insignificant, a two-loop calculation would be certainly welcome; this is further necessitated by the fact that the one-loop contributions for the real parts of the diagonal matrix elements of the conversion factors do not sufficiently stabilize for large values of z.

\clearpage

\section{Conclusions and Follow-up work} 

In this paper, we have presented the one-loop calculation, in dimensional regularization, of the renormalization factors for nonlocal quark operators, including a straight Wilson line, which are involved in the definition of quasi-PDFs. The novel aspect of this work is the presence of nonzero quark masses in our computations, which results in mixing among these operators, both in the continuum and on the lattice. 

\bigskip

The operator mixing, observed in Ref.\cite{Constantinou:2017sej} for massless fermions on the lattice, is extended into more operator pairs for massive fermions. More precisely, for operators with equal masses of external quark fields, the mixing pairs are the same as those of massless fermions; i.e., the unpolarized quasi-PDF in direction $\mu$ (parallel to the Wilson line) mixes with the twist-3 scalar operator, and the helicity quasi-PDF in direction $\nu$ (perpendicular to $\mu$) mixes with the transversity quasi-PDF in directions perpendicular to $\mu$ and $\nu$. However, for operators with different masses of external quark fields, there are additional pairs: the helicity quasi-PDF in direction $\mu$ mixes with the pseudoscalar operator, and the unpolarized quasi-PDF in direction $\nu$ mixes with the transversity quasi-PDF in the $\mu$ and $\nu$ directions. Thus, before matching to the physical massive PDFs, one must eliminate the mixing nonperturbatively. To this end, we extend the RI$'$ scheme suggested in Ref.\cite{Constantinou:2017sej} including the additional mixing pairs. 

\bigskip

To convert the nonperturbative RI$'$ estimates of renormalization factors to the $\MSbar$ scheme, we have calculated the one-loop conversion factors between the two schemes in DR for massive quarks. Because of the operator-pair mixing in the continuum, the conversion factors are generally nondiagonal 2$\times$2 matrices. Comparing with the massless case, the impact of quark masses on the conversion factors becomes significant for values near or greater than the strange quark mass.            

\bigskip
 
A natural continuation of the present work is the two-loop calculation of the renormalization factors in DR, as well as the conversion factors between the RI$'$ and $\MSbar$ schemes. According to Ref.\cite{Alexandrou:2017huk}, two-loop corrections of the conversion factors are expected to suppress the unphysical observed feature of the negative real part of the nonperturbative renormalized matrix elements for large Wilson-line lengths. A by-product of this two-loop calculation is the anomalous dimension of the operators to next order in $g^2$, which can be found in Refs.\cite{Ji:1991pr,Broadhurst:1991fz,Chetyrkin:2003vi}; this is useful for improving the method for eliminating the linear divergences, nonperturbatively (see Ref.\cite{Constantinou:2017sej}). 

\bigskip

Another extension of our work is the perturbative study of Wilson-line operators with more composite Wilson lines, such as ``staples''. Here, the appearance of an additional special direction (specifying the plane on which the staple lies) may give us further operator-mixing patterns. Thus, this perturbative investigation can be a guidance to the development of a nonperturbative renormalization prescription for eliminating mixing and linear divergences in these operators as well. Such staple operators are involved in the definition of TMDs, which are currently under investigation for the nucleon and pion in lattice QCD \cite{Engelhardt:2015xja, Radyushkin:2016hsy, Yoon:2017qzo}. Our findings will be presented separately in a future publication.   

\clearpage
\centerline{{\bf\large{{\bf{Acknowledgements}}}}}
\bigskip
H.P. would like to thank M. Constantinou for useful discussions. G.S. acknowledges financial support by the University of Cyprus, within the framework of Ph.D. student scholarships.

\appendix
\section{The integration method}

In this Appendix, we describe the method that we used to evaluate the D-dimensional momentum-loop integrals, appearing in the calculation of the Feynman diagrams of Fig. \ref{Fig.FeynmanDiagrams}. First, we introduce Feynman parameters. Second, we perform the standard integrations over the $(D - 1)$ directions perpendicular to the Wilson line (see, e.g., Ref. \cite{tHooft:1973wag}). Next, we perform the remaining nontrivial integration over the parallel direction, which has an exponential $z$ dependence. This procedure gives the following formulae, in terms of modified Bessel functions of the second kind, $K_\nu$: 
\be 
A (\alpha) = \int \frac{d^D p}{(2 \pi)^D} \frac{e^{i p_\mu z}}{{(p^2 + 2 \ k \cdot p + m^2)}^{\alpha}} = \frac{2^{1-\alpha-D/2} \ {\vert z \vert}^{\alpha - D/2} \ e^{-i k_{\mu} z}}{\pi^{D/2} \ \Gamma (\alpha) \ (m^2 - k^2)^{\alpha/2 - D/4}} \ K_{-\alpha + D/2} (\sqrt{m^2-k^2} \ \vert z \vert),
\ee
\be 
\int \frac{d^D p}{(2 \pi)^D} \frac{e^{i p_\mu z} \ p_{\nu_1} \cdots p_{\nu_n}}{{(p^2 + 2 \ k \cdot p + m^2)}^{\alpha}} = \frac{{(-1)}^n \ \Gamma (\alpha - n)}{2^n \ \Gamma (\alpha)} \frac{\partial}{\partial k_{\nu_1}} \cdots \frac{\partial}{\partial k_{\nu_n}} A (\alpha - n).
\ee

\bigskip

After the momentum integrations, we perform Laurent expansion in $\varepsilon$, keeping terms up to $\mathcal{O} (\varepsilon^0)$. In this step, we have to be careful when interchanging the integration over Feynman parameters with the limit of a vanishing regulator ($\varepsilon \rightarrow 0$). In the massive case, studied in the present paper, the interchange is permissible; however this interchange is not generally valid, as is exemplified by the following term stemming from diagram 1, in the massless case\footnote{Diagram 1 is actually UV convergent; however, in order to avoid spurious IR divergences, it is convenient to evaluate it in $D>4$ dimensions ($\varepsilon < 0$) and take the limit $\varepsilon \rightarrow 0^-$ in the end.}:
\be 
B(\varepsilon) = \int_0^1 dx \ \frac{\exp (\imag q_\mu z x) \ q^2 x^2 \left| z \right|^{1 + \varepsilon} \ \varepsilon}{\Big(q^2 \ x \ (1-x) \Big)^{(1+\varepsilon)/2}} \ K_{1+\varepsilon} (\sqrt{q^2 x (1-x)} \left| z \right|).
\label{integral}
\ee
A naive limit $\varepsilon \rightarrow 0^-$ of this term would simply give 0, due to the multiplicative factor of $\varepsilon$. However, this is incorrect, given the existence of a pole at x = 1. Expanding the integrand of Eq. \eqref{integral} into a power series of $(1-x)$:
\be
K_{1+\varepsilon} (\sqrt{q^2 x (1-x)} \left| z \right|) = \frac{1}{2} \ \Gamma (1+\varepsilon) \frac{(\sqrt{q^2 x (1-x)} \left| z \right|)^{-1-\varepsilon}}{2^{-1-\varepsilon}} + \mathcal{O} \Big( (1-x)^{(1+\varepsilon)/2} \Big),
\ee
\be 
\exp(\imag q_\mu z x) = \exp(\imag q_\mu z) + \mathcal{O} (1-x),
\ee
we isolate the pole:
\be 
\int_0^1 dx \ \Big[ 2^{\varepsilon} \ \varepsilon \ \Gamma (1+\varepsilon) \ \frac{\exp (\imag q_\mu z)}{(q^2)^\varepsilon (1-x)^{1 + \varepsilon}} + \mathcal{O} \Big( (1-x)^{(1+\varepsilon)/2} \Big) \Big].
\label{integral2}
\ee
The terms of order $\mathcal{O} \Big((1-x)^{(1+\varepsilon)/2} \Big)$ are integrable in the limit $\varepsilon \rightarrow 0^-$, and thus they give 0. In the leading term of Eq. \eqref{integral2}, we must perform the Feynman parameter integral first, and after that, we take the limit $\varepsilon \to 0^-$. Then, a finite but nonzero result remains:
\be
\lim_{\varepsilon \rightarrow 0^-} B(\varepsilon) = - \exp (\imag q_\mu z).
\ee
Therefore, the naive interchange of limit and integration sets a contribution erroneously to zero. To avoid such errors, we use a subtraction of the form:
\be  
\lim_{\varepsilon \to 0} \int dx \ I(\varepsilon, x) = \int dx \lim_{\varepsilon \to 0} \Big( I(\varepsilon,x) - I_1 (\varepsilon,x) \Big) + \lim_{\varepsilon \to 0} \int dx \ I_1 (\varepsilon,x),
\ee
where $I(\varepsilon,x)$ is a term of the original expression and $I_1(\varepsilon,x)$ denotes the leading terms of $I(\varepsilon,x)$ in a power series expansion with respect to $(x-x_i)$ about all singular points $x_i$; here, $x$ denotes Feynman parameters and/or $\zeta$ variables stemming from the definition of $\mathcal{O}_\Gamma$. Such a subtraction must also be applied when we take the massless limit of our results, $m \to 0$, for the same reasons. 

\bigskip

The final expression depends on the Feynman parameter integrals and/or the integrals stemming from the definition of $\mathcal{O}_\Gamma$; these can be integrated numerically for all values of $q$, $z$, and quark masses used in simulations. 

\section{Table of Feynman parameter Integrals}

In this Appendix, we present a table of Feynman parameter integrals, which appear in the expressions of our results. They do not have a closed analytic form, but they are convergent and can be computed numerically. We can classify them into three types of integrals: \\ 
1. $f_1$ - $f_3$: integrals over the Feynman parameter $x$ \\
2. $g_1$ - $g_5$: double integrals over the Feynman parameter $x$ and variable $\zeta$ (the location of gluon fields along the Wilson line) \\
3. $h_1$ - $h_8$: double integrals over the Feynman parameters $x_1$ and $x_2$. \\
These integrals are functions of the external momentum 4-vector $q_\nu$, the Wilson-line length $z$, and the external quark masses $m_1$ and/or $m_2$. Also, they involve a modified Bessel function of the second kind, $K_0$ or $K_1$. For the sake of brevity, we use the following notation: $s \equiv {\Big(q^2 \left(1-x\right) x + m^2 x\Big)}^{1/2}$ and \ \ $t \equiv {\Big( q^2 \left(1-x_1 - x_2\right) \left(x_1 + x_2\right) + m_1^2 \ x_1 + m_2^2 \ x_2 \Big)}^{1/2}$,

\begin{align}
f_1 \left(q,z,m\right) &= \int_0^1 dx \ \exp{\left(-\imag q_{\mu} x z \right)} \ K_0\left( \left| z\right| s \right), \label{f1}\\
f_2 \left(q,z,m\right) &= \int_0^1 dx \ \exp{\left(-\imag q_{\mu} x z \right)} \ K_0\left( \left| z\right| s \right) \ (1-x), \\
f_3 \left(q,z,m\right) &= \int_0^1 dx \ \exp{\left(-\imag q_{\mu} x z \right)} \ K_0\left( \left| z\right| s \right) \ (1 - x) \ \frac{x^2}{{s}^2},
\end{align}
\begin{align}
g_1 \left(q,z,m\right) &= \int_0^1 dx \ \int_0^z d\zeta \ \exp{\left(-\imag q_{\mu} x \zeta \right)} \ K_0\left( \left| \zeta \right| s \right), \\
g_2 \left(q,z,m\right) &= \int_0^1 dx \ \int_0^z d\zeta \ \exp{\left(-\imag q_{\mu} x \zeta \right)} \ K_0\left( \left| \zeta \right| s \right) \ x, \\
g_3 \left(q,z,m\right) &= \int_0^1 dx \ \int_0^z d\zeta \ \exp{\left(-\imag q_{\mu} x \zeta \right)} \ K_0\left( \left| \zeta \right| s \right) \ (1 - x) \ \frac{x^2}{s^2}, \\
g_4 \left(q,z,m\right) &= \int_0^1 dx \ \int_0^z d\zeta \ \exp{\left(-\imag q_{\mu} x \zeta \right)} \ K_0\left( \left| \zeta \right| s \right) \ (1 - x) \ \frac{x^2}{s^2} \ \zeta, \\
g_5 \left(q,z,m\right) &= \int_0^1 dx \ \int_0^z d\zeta \ \exp{\left(-\imag q_{\mu} x \zeta \right)} \ K_0\left( \left| \zeta \right| s \right) \ (1 - x) \ \frac{x^3}{s^2} \ \zeta,
\end{align}
\begin{align}
h_1 \left(q,z,m_1,m_2\right) &= \int_0^1 dx_1 \ \int_0^{1 - x_1} \hspace{-0.25cm}dx_2 \ \exp{\left(-\imag q_{\mu} (x_1 + x_2) z \right)} \ K_0\left( \left| z\right| t \right), \\
h_2 \left(q,z,m_1,m_2\right) &= \int_0^1 dx_1 \ \int_0^{1 - x_1} \hspace{-0.25cm}dx_2 \ \exp{\left(-\imag q_{\mu} (x_1 + x_2) z \right)} \ K_0\left( \left| z\right| t \right) \ (1 - x_1 - x_2), \\
h_3 \left(q,z,m_1,m_2\right) &= \int_0^1 dx_1 \ \int_0^{1 - x_1} \hspace{-0.25cm}dx_2 \ \exp{\left(-\imag q_{\mu} (x_1 + x_2) z \right)} \ K_0\left( \left| z\right| t \right) \ (1 - x_1 - x_2) \cdot \nonumber \\
& \qquad \qquad \qquad \qquad \quad \ \frac{(x_1 + x_2)^2}{t^2}, \\
h_4 \left(q,z,m_1,m_2\right) &= \int_0^1 dx_1 \ \int_0^{1 - x_1} \hspace{-0.25cm}dx_2 \ \exp{\left(-\imag q_{\mu} (x_1 + x_2) z \right)} \ K_1\left( \left| z\right| t \right) \ t, \\
h_5 \left(q,z,m_1,m_2\right) &= \int_0^1 dx_1 \ \int_0^{1 - x_1} \hspace{-0.25cm}dx_2 \ \exp{\left(-\imag q_{\mu} (x_1 + x_2) z \right)} \ K_1\left( \left| z\right| t \right) \ \frac{1}{t}, \\
h_6 \left(q,z,m_1,m_2\right) &= \int_0^1 dx_1 \ \int_0^{1 - x_1} \hspace{-0.25cm}dx_2 \ \exp{\left(-\imag q_{\mu} (x_1 + x_2) z \right)} \ K_1\left( \left| z\right| t \right) \ \frac{(x_1 + x_2)}{t}, \\
h_7 \left(q,z,m_1,m_2\right) &= \int_0^1 dx_1 \ \int_0^{1 - x_1} \hspace{-0.25cm}dx_2 \ \exp{\left(-\imag q_{\mu} (x_1 + x_2) z \right)} \ K_1\left( \left| z\right| t \right) \ \frac{(1 - x_1 - x_2)^2}{t}, \\
h_8 \left(q,z,m_1,m_2\right) &= \int_0^1 dx_1 \ \int_0^{1 - x_1} \hspace{-0.25cm}dx_2 \ \exp{\left(-\imag q_{\mu} (x_1 + x_2) z \right)} \ K_1\left( \left| z\right| t \right) \ (1 - x_1 - x_2) \cdot \nonumber \\
& \qquad \qquad \qquad \qquad \quad \ \frac{(x_1 + x_2)^2}{t^3}. 
\label{h8}
\end{align}

\vspace*{-0.1cm}
   
\bibliographystyle{elsarticle-num}                     
\bibliography{references}

\begin{thebibliography}{10}
\expandafter\ifx\csname url\endcsname\relax
  \def\url#1{\texttt{#1}}\fi
\expandafter\ifx\csname urlprefix\endcsname\relax\def\urlprefix{URL }\fi
\expandafter\ifx\csname href\endcsname\relax
  \def\href#1#2{#2} \def\path#1{#1}\fi

\bibitem{Ji:2013dva}
X.~Ji, {Parton Physics on a Euclidean Lattice}, Phys. Rev. Lett. 110 (2013)
  262002.
\newblock \href {http://arxiv.org/abs/1305.1539} {\path{arXiv:1305.1539}},
  \href {http://dx.doi.org/10.1103/PhysRevLett.110.262002}
  {\path{doi:10.1103/PhysRevLett.110.262002}}.

\bibitem{Ji:2014gla}
X.~Ji, {Parton Physics from Large-Momentum Effective Field Theory}, Sci. China
  Phys. Mech. Astron. 57 (2014) 1407--1412.
\newblock \href {http://arxiv.org/abs/1404.6680} {\path{arXiv:1404.6680}},
  \href {http://dx.doi.org/10.1007/s11433-014-5492-3}
  {\path{doi:10.1007/s11433-014-5492-3}}.

\bibitem{Xiong:2013bka}
X.~Xiong, X.~Ji, J.-H. Zhang, Y.~Zhao, {One-loop matching for parton
  distributions: Nonsinglet case}, Phys. Rev. D90~(1) (2014) 014051.
\newblock \href {http://arxiv.org/abs/1310.7471} {\path{arXiv:1310.7471}},
  \href {http://dx.doi.org/10.1103/PhysRevD.90.014051}
  {\path{doi:10.1103/PhysRevD.90.014051}}.

\bibitem{Lin:2014zya}
H.-W. Lin, J.-W. Chen, S.~D. Cohen, X.~Ji, {Flavor Structure of the Nucleon Sea
  from Lattice QCD}, Phys. Rev. D91 (2015) 054510.
\newblock \href {http://arxiv.org/abs/1402.1462} {\path{arXiv:1402.1462}},
  \href {http://dx.doi.org/10.1103/PhysRevD.91.054510}
  {\path{doi:10.1103/PhysRevD.91.054510}}.

\bibitem{Alexandrou:2014pna}
C.~Alexandrou, K.~Cichy, V.~Drach, E.~Garcia-Ramos, K.~Hadjiyiannakou,
  K.~Jansen, F.~Steffens, C.~Wiese, {First results with twisted mass fermions
  towards the computation of parton distribution functions on the lattice}, PoS
  LATTICE2014 (2014) 135.
\newblock \href {http://arxiv.org/abs/1411.0891} {\path{arXiv:1411.0891}},
  \href {http://dx.doi.org/10.22323/1.214.0135}
  {\path{doi:10.22323/1.214.0135}}.

\bibitem{Gamberg:2014zwa}
L.~Gamberg, Z.-B. Kang, I.~Vitev, H.~Xing, {Quasi-parton distribution
  functions: a study in the diquark spectator model}, Phys. Lett. B743 (2015)
  112--120.
\newblock \href {http://arxiv.org/abs/1412.3401} {\path{arXiv:1412.3401}},
  \href {http://dx.doi.org/10.1016/j.physletb.2015.02.021}
  {\path{doi:10.1016/j.physletb.2015.02.021}}.

\bibitem{Alexandrou:2015rja}
C.~Alexandrou, K.~Cichy, V.~Drach, E.~Garcia-Ramos, K.~Hadjiyiannakou,
  K.~Jansen, F.~Steffens, C.~Wiese, {Lattice calculation of parton
  distributions}, Phys. Rev. D92 (2015) 014502.
\newblock \href {http://arxiv.org/abs/1504.07455} {\path{arXiv:1504.07455}},
  \href {http://dx.doi.org/10.1103/PhysRevD.92.014502}
  {\path{doi:10.1103/PhysRevD.92.014502}}.

\bibitem{Gamberg:2015opc}
I.~Vitev, L.~Gamberg, Z.~Kang, H.~Xing, {A Study of Quasi-parton Distribution
  Functions in the Diquark Spectator Model}, PoS QCDEV2015 (2015) 045.
\newblock \href {http://arxiv.org/abs/1511.05242} {\path{arXiv:1511.05242}},
  \href {http://dx.doi.org/10.22323/1.249.0045}
  {\path{doi:10.22323/1.249.0045}}.

\bibitem{Chen:2016utp}
J.-W. Chen, S.~D. Cohen, X.~Ji, H.-W. Lin, J.-H. Zhang, {Nucleon Helicity and
  Transversity Parton Distributions from Lattice QCD}, Nucl. Phys. B911 (2016)
  246--273.
\newblock \href {http://arxiv.org/abs/1603.06664} {\path{arXiv:1603.06664}},
  \href {http://dx.doi.org/10.1016/j.nuclphysb.2016.07.033}
  {\path{doi:10.1016/j.nuclphysb.2016.07.033}}.

\bibitem{Bacchetta:2016zjm}
A.~Bacchetta, M.~Radici, B.~Pasquini, X.~Xiong, {Reconstructing parton
  densities at large fractional momenta}, Phys. Rev. D95~(1) (2017) 014036.
\newblock \href {http://arxiv.org/abs/1608.07638} {\path{arXiv:1608.07638}},
  \href {http://dx.doi.org/10.1103/PhysRevD.95.014036}
  {\path{doi:10.1103/PhysRevD.95.014036}}.

\bibitem{Alexandrou:2016bud}
C.~Alexandrou, K.~Cichy, K.~Hadjiyiannakou, K.~Jansen, F.~Steffens, C.~Wiese,
  {A Lattice Calculation of Parton Distributions}, PoS DIS2016 (2016) 042.
\newblock \href {http://arxiv.org/abs/1609.00172} {\path{arXiv:1609.00172}},
  \href {http://dx.doi.org/10.22323/1.265.0042}
  {\path{doi:10.22323/1.265.0042}}.

\bibitem{Alexandrou:2016jqi}
C.~Alexandrou, K.~Cichy, M.~Constantinou, K.~Hadjiyiannakou, K.~Jansen,
  F.~Steffens, C.~Wiese, {Updated Lattice Results for Parton Distributions},
  Phys. Rev. D96~(1) (2017) 014513.
\newblock \href {http://arxiv.org/abs/1610.03689} {\path{arXiv:1610.03689}},
  \href {http://dx.doi.org/10.1103/PhysRevD.96.014513}
  {\path{doi:10.1103/PhysRevD.96.014513}}.

\bibitem{Alexandrou:2016eyt}
C.~Alexandrou, K.~Cichy, M.~Constantinou, K.~Hadjiyiannakou, K.~Jansen,
  F.~Steffens, C.~Wiese, {Parton Distributions from Lattice QCD with Momentum
  Smearing}, PoS LATTICE2016 (2016) 151.
\newblock \href {http://arxiv.org/abs/1612.08728} {\path{arXiv:1612.08728}},
  \href {http://dx.doi.org/10.22323/1.256.0151}
  {\path{doi:10.22323/1.256.0151}}.

\bibitem{Carlson:2017gpk}
C.~E. Carlson, M.~Freid, {Lattice corrections to the quark quasidistribution at
  one-loop}, Phys. Rev. D95~(9) (2017) 094504.
\newblock \href {http://arxiv.org/abs/1702.05775} {\path{arXiv:1702.05775}},
  \href {http://dx.doi.org/10.1103/PhysRevD.95.094504}
  {\path{doi:10.1103/PhysRevD.95.094504}}.

\bibitem{Xiong:2017jtn}
X.~Xiong, T.~Luu, U.-G. Meißner, {Quasi-Parton Distribution Function in
  Lattice Perturbation Theory \ }\href {http://arxiv.org/abs/1705.00246}
  {\path{arXiv:1705.00246}}.

\bibitem{Constantinou:2017sej}
M.~Constantinou, H.~Panagopoulos, {Perturbative renormalization of quasi-parton
  distribution functions}, Phys. Rev. D96~(5) (2017) 054506.
\newblock \href {http://arxiv.org/abs/1705.11193} {\path{arXiv:1705.11193}},
  \href {http://dx.doi.org/10.1103/PhysRevD.96.054506}
  {\path{doi:10.1103/PhysRevD.96.054506}}.

\bibitem{Ma:2014jla}
Y.-Q. Ma, J.-W. Qiu, {Extracting Parton Distribution Functions from Lattice QCD
  Calculations \ }\href {http://arxiv.org/abs/1404.6860}
  {\path{arXiv:1404.6860}}.

\bibitem{Ma:2014jga}
Y.-Q. Ma, J.-W. Qiu, {QCD Factorization and PDFs from Lattice QCD Calculation},
  Int. J. Mod. Phys. Conf. Ser. 37 (2015) 1560041.
\newblock \href {http://arxiv.org/abs/1412.2688} {\path{arXiv:1412.2688}},
  \href {http://dx.doi.org/10.1142/S2010194515600411}
  {\path{doi:10.1142/S2010194515600411}}.

\bibitem{Li:2016amo}
H.-n. Li, {Nondipolar Wilson links for quasiparton distribution functions},
  Phys. Rev. D94~(7) (2016) 074036.
\newblock \href {http://arxiv.org/abs/1602.07575} {\path{arXiv:1602.07575}},
  \href {http://dx.doi.org/10.1103/PhysRevD.94.074036}
  {\path{doi:10.1103/PhysRevD.94.074036}}.

\bibitem{Chen:2016fxx}
J.-W. Chen, X.~Ji, J.-H. Zhang, {Improved quasi parton distribution through
  Wilson line renormalization}, Nucl. Phys. B915 (2017) 1--9.
\newblock \href {http://arxiv.org/abs/1609.08102} {\path{arXiv:1609.08102}},
  \href {http://dx.doi.org/10.1016/j.nuclphysb.2016.12.004}
  {\path{doi:10.1016/j.nuclphysb.2016.12.004}}.

\bibitem{Wang:2017qyg}
W.~Wang, S.~Zhao, R.~Zhu, {Gluon quasidistribution function at one loop}, Eur.
  Phys. J. C78~(2) (2018) 147.
\newblock \href {http://arxiv.org/abs/1708.02458} {\path{arXiv:1708.02458}},
  \href {http://dx.doi.org/10.1140/epjc/s10052-018-5617-3}
  {\path{doi:10.1140/epjc/s10052-018-5617-3}}.

\bibitem{Izubuchi:2018srq}
T.~Izubuchi, X.~Ji, L.~Jin, I.~W. Stewart, Y.~Zhao, {Factorization Theorem
  Relating Euclidean and Light-Cone Parton Distributions \ }\href
  {http://arxiv.org/abs/1801.03917} {\path{arXiv:1801.03917}}.

\bibitem{Briceno:2017cpo}
R.~A. Briceño, M.~T. Hansen, C.~J. Monahan, {Role of the Euclidean signature
  in lattice calculations of quasi-distributions and other nonlocal matrix
  elements}, Phys. Rev. D96~(1) (2017) 014502.
\newblock \href {http://arxiv.org/abs/1703.06072} {\path{arXiv:1703.06072}},
  \href {http://dx.doi.org/10.1103/PhysRevD.96.014502}
  {\path{doi:10.1103/PhysRevD.96.014502}}.

\bibitem{Rossi:2017muf}
G.~C. Rossi, M.~Testa, {Note on lattice regularization and equal-time
  correlators for parton distribution functions}, Phys. Rev. D96~(1) (2017)
  014507.
\newblock \href {http://arxiv.org/abs/1706.04428} {\path{arXiv:1706.04428}},
  \href {http://dx.doi.org/10.1103/PhysRevD.96.014507}
  {\path{doi:10.1103/PhysRevD.96.014507}}.

\bibitem{Ji:2017rah}
X.~Ji, J.-H. Zhang, Y.~Zhao, {More On Large-Momentum Effective Theory Approach
  to Parton Physics}, Nucl. Phys. B924 (2017) 366--376.
\newblock \href {http://arxiv.org/abs/1706.07416} {\path{arXiv:1706.07416}},
  \href {http://dx.doi.org/10.1016/j.nuclphysb.2017.09.001}
  {\path{doi:10.1016/j.nuclphysb.2017.09.001}}.

\bibitem{Ji:2014hxa}
X.~Ji, P.~Sun, X.~Xiong, F.~Yuan, {Soft factor subtraction and transverse
  momentum dependent parton distributions on the lattice}, Phys. Rev. D91
  (2015) 074009.
\newblock \href {http://arxiv.org/abs/1405.7640} {\path{arXiv:1405.7640}},
  \href {http://dx.doi.org/10.1103/PhysRevD.91.074009}
  {\path{doi:10.1103/PhysRevD.91.074009}}.

\bibitem{Engelhardt:2015xja}
M.~Engelhardt, P.~Hägler, B.~Musch, J.~Negele, A.~Schäfer, {Lattice QCD study
  of the Boer-Mulders effect in a pion}, Phys. Rev. D93~(5) (2016) 054501.
\newblock \href {http://arxiv.org/abs/1506.07826} {\path{arXiv:1506.07826}},
  \href {http://dx.doi.org/10.1103/PhysRevD.93.054501}
  {\path{doi:10.1103/PhysRevD.93.054501}}.

\bibitem{Radyushkin:2016hsy}
A.~Radyushkin, {Nonperturbative Evolution of Parton Quasi-Distributions}, Phys.
  Lett. B767 (2017) 314--320.
\newblock \href {http://arxiv.org/abs/1612.05170} {\path{arXiv:1612.05170}},
  \href {http://dx.doi.org/10.1016/j.physletb.2017.02.019}
  {\path{doi:10.1016/j.physletb.2017.02.019}}.

\bibitem{Radyushkin:2017ffo}
A.~Radyushkin, {Target Mass Effects in Parton Quasi-Distributions}, Phys. Lett.
  B770 (2017) 514--522.
\newblock \href {http://arxiv.org/abs/1702.01726} {\path{arXiv:1702.01726}},
  \href {http://dx.doi.org/10.1016/j.physletb.2017.05.024}
  {\path{doi:10.1016/j.physletb.2017.05.024}}.

\bibitem{Yoon:2017qzo}
B.~Yoon, M.~Engelhardt, R.~Gupta, T.~Bhattacharya, J.~R. Green, B.~U. Musch,
  J.~W. Negele, A.~V. Pochinsky, A.~Schäfer, S.~N. Syritsyn, {Nucleon
  Transverse Momentum-dependent Parton Distributions in Lattice QCD:
  Renormalization Patterns and Discretization Effects}, Phys. Rev. D96~(9)
  (2017) 094508.
\newblock \href {http://arxiv.org/abs/1706.03406} {\path{arXiv:1706.03406}},
  \href {http://dx.doi.org/10.1103/PhysRevD.96.094508}
  {\path{doi:10.1103/PhysRevD.96.094508}}.

\bibitem{Broniowski:2017gfp}
W.~Broniowski, E.~Ruiz~Arriola, {Partonic quasidistributions of the proton and
  pion from transverse-momentum distributions}, Phys. Rev. D97~(3) (2018)
  034031.
\newblock \href {http://arxiv.org/abs/1711.03377} {\path{arXiv:1711.03377}},
  \href {http://dx.doi.org/10.1103/PhysRevD.97.034031}
  {\path{doi:10.1103/PhysRevD.97.034031}}.

\bibitem{Ji:2018hvs}
X.~Ji, L.-C. Jin, F.~Yuan, J.-H. Zhang, Y.~Zhao, {Transverse Momentum Dependent
  Quasi-Parton-Distributions \ }\href {http://arxiv.org/abs/1801.05930}
  {\path{arXiv:1801.05930}}.

\bibitem{Ji:2015qla}
X.~Ji, A.~Schäfer, X.~Xiong, J.-H. Zhang, {One-Loop Matching for Generalized
  Parton Distributions}, Phys. Rev. D92 (2015) 014039.
\newblock \href {http://arxiv.org/abs/1506.00248} {\path{arXiv:1506.00248}},
  \href {http://dx.doi.org/10.1103/PhysRevD.92.014039}
  {\path{doi:10.1103/PhysRevD.92.014039}}.

\bibitem{Xiong:2015nua}
X.~Xiong, J.-H. Zhang, {One-loop matching for transversity generalized parton
  distribution}, Phys. Rev. D92~(5) (2015) 054037.
\newblock \href {http://arxiv.org/abs/1509.08016} {\path{arXiv:1509.08016}},
  \href {http://dx.doi.org/10.1103/PhysRevD.92.054037}
  {\path{doi:10.1103/PhysRevD.92.054037}}.

\bibitem{Jia:2015pxx}
Y.~Jia, X.~Xiong, {Quasi-distribution amplitude of heavy quarkonia}, Phys. Rev.
  D94~(9) (2016) 094005.
\newblock \href {http://arxiv.org/abs/1511.04430} {\path{arXiv:1511.04430}},
  \href {http://dx.doi.org/10.1103/PhysRevD.94.094005}
  {\path{doi:10.1103/PhysRevD.94.094005}}.

\bibitem{Radyushkin:2017gjd}
A.~V. Radyushkin, {Pion Distribution Amplitude and Quasi-Distributions}, Phys.
  Rev. D95~(5) (2017) 056020.
\newblock \href {http://arxiv.org/abs/1701.02688} {\path{arXiv:1701.02688}},
  \href {http://dx.doi.org/10.1103/PhysRevD.95.056020}
  {\path{doi:10.1103/PhysRevD.95.056020}}.

\bibitem{Zhang:2017bzy}
J.-H. Zhang, J.-W. Chen, X.~Ji, L.~Jin, H.-W. Lin, {Pion Distribution Amplitude
  from Lattice QCD}, Phys. Rev. D95~(9) (2017) 094514.
\newblock \href {http://arxiv.org/abs/1702.00008} {\path{arXiv:1702.00008}},
  \href {http://dx.doi.org/10.1103/PhysRevD.95.094514}
  {\path{doi:10.1103/PhysRevD.95.094514}}.

\bibitem{Broniowski:2017wbr}
W.~Broniowski, E.~Ruiz~Arriola, {Nonperturbative partonic quasidistributions of
  the pion from chiral quark models}, Phys. Lett. B773 (2017) 385--390.
\newblock \href {http://arxiv.org/abs/1707.09588} {\path{arXiv:1707.09588}},
  \href {http://dx.doi.org/10.1016/j.physletb.2017.08.055}
  {\path{doi:10.1016/j.physletb.2017.08.055}}.

\bibitem{Chen:2017gck}
J.-W. Chen, L.~Jin, H.-W. Lin, A.~Schäfer, P.~Sun, Y.-B. Yang, J.-H. Zhang,
  R.~Zhang, Y.~Zhao, {Kaon Distribution Amplitude from Lattice QCD and the
  Flavor SU(3) Symmetry \ }\href {http://arxiv.org/abs/1712.10025}
  {\path{arXiv:1712.10025}}.

\bibitem{Ji:2014lra}
X.~Ji, J.-H. Zhang, Y.~Zhao, {Justifying the Naive Partonic Sum Rule for Proton
  Spin}, Phys. Lett. B743 (2015) 180--183.
\newblock \href {http://arxiv.org/abs/1409.6329} {\path{arXiv:1409.6329}},
  \href {http://dx.doi.org/10.1016/j.physletb.2015.02.054}
  {\path{doi:10.1016/j.physletb.2015.02.054}}.

\bibitem{Lin:2017snn}
H.-W. Lin, et~al., {Parton distributions and lattice QCD calculations: a
  community white paper}, Prog. Part. Nucl. Phys. 100 (2018) 107--160.
\newblock \href {http://arxiv.org/abs/1711.07916} {\path{arXiv:1711.07916}},
  \href {http://dx.doi.org/10.1016/j.ppnp.2018.01.007}
  {\path{doi:10.1016/j.ppnp.2018.01.007}}.

\bibitem{Ji:2015jwa}
X.~Ji, J.-H. Zhang, {Renormalization of quasi-parton distribution}, Phys. Rev.
  D92 (2015) 034006.
\newblock \href {http://arxiv.org/abs/1505.07699} {\path{arXiv:1505.07699}},
  \href {http://dx.doi.org/10.1103/PhysRevD.92.034006}
  {\path{doi:10.1103/PhysRevD.92.034006}}.

\bibitem{Ishikawa:2017faj}
T.~Ishikawa, Y.-Q. Ma, J.-W. Qiu, S.~Yoshida, {Renormalizability of
  quasi-parton distribution functions}, Phys. Rev. D96~(9) (2017) 094019.
\newblock \href {http://arxiv.org/abs/1707.03107} {\path{arXiv:1707.03107}},
  \href {http://dx.doi.org/10.1103/PhysRevD.96.094019}
  {\path{doi:10.1103/PhysRevD.96.094019}}.

\bibitem{Dotsenko:1979wb}
V.~S. Dotsenko, S.~N. Vergeles, {Renormalizability of Phase Factors in the
  Nonabelian Gauge Theory}, Nucl. Phys. B169 (1980) 527--546.
\newblock \href {http://dx.doi.org/10.1016/0550-3213(80)90103-0}
  {\path{doi:10.1016/0550-3213(80)90103-0}}.

\bibitem{Brandt:1981kf}
R.~A. Brandt, F.~Neri, M.-a. Sato, {Renormalization of Loop Functions for All
  Loops}, Phys. Rev. D24 (1981) 879.
\newblock \href {http://dx.doi.org/10.1103/PhysRevD.24.879}
  {\path{doi:10.1103/PhysRevD.24.879}}.

\bibitem{Constantinou:2017fiy}
M.~Constantinou, H.~Panagopoulos, {Perturbative Renormalization of Wilson line
  operators}, EPJ Web of Conferences 175 (Lattice 2017) (2018) 06025.
\newblock \href {http://arxiv.org/abs/1711.00543} {\path{arXiv:1711.00543}},
  \href {http://dx.doi.org/10.1051/epjconf/201817506025}
  {\path{doi:10.1051/epjconf/201817506025}}.

\bibitem{Ishikawa:2016znu}
T.~Ishikawa, Y.-Q. Ma, J.-W. Qiu, S.~Yoshida, {Practical quasi parton
  distribution functions \ }\href {http://arxiv.org/abs/1609.02018}
  {\path{arXiv:1609.02018}}.

\bibitem{Ishikawa:2017jtf}
T.~Ishikawa, Y.-Q. Ma, J.-W. Qiu, S.~Yoshida, {Matching issue in quasi parton
  distribution approach}, PoS LATTICE2016 (2016) 177.
\newblock \href {http://arxiv.org/abs/1703.08699} {\path{arXiv:1703.08699}},
  \href {http://dx.doi.org/10.22323/1.256.0177}
  {\path{doi:10.22323/1.256.0177}}.

\bibitem{Monahan:2015lha}
C.~Monahan, K.~Orginos, {Locally smeared operator product expansions in scalar
  field theory}, Phys. Rev. D91~(7) (2015) 074513.
\newblock \href {http://arxiv.org/abs/1501.05348} {\path{arXiv:1501.05348}},
  \href {http://dx.doi.org/10.1103/PhysRevD.91.074513}
  {\path{doi:10.1103/PhysRevD.91.074513}}.

\bibitem{Monahan:2016bvm}
C.~Monahan, K.~Orginos, {Quasi parton distributions and the gradient flow},
  JHEP 03 (2017) 116.
\newblock \href {http://arxiv.org/abs/1612.01584} {\path{arXiv:1612.01584}},
  \href {http://dx.doi.org/10.1007/JHEP03(2017)116}
  {\path{doi:10.1007/JHEP03(2017)116}}.

\bibitem{Monahan:2017hpu}
C.~Monahan, {Smeared quasidistributions in perturbation theory}, Phys. Rev.
  D97~(5) (2018) 054507.
\newblock \href {http://arxiv.org/abs/1710.04607} {\path{arXiv:1710.04607}},
  \href {http://dx.doi.org/10.1103/PhysRevD.97.054507}
  {\path{doi:10.1103/PhysRevD.97.054507}}.

\bibitem{Dorn:1986dt}
H.~Dorn, {Renormalization of Path Ordered Phase Factors and Related Hadron
  Operators in Gauge Field Theories}, Fortsch. Phys. 34 (1986) 11--56.
\newblock \href {http://dx.doi.org/10.1002/prop.19860340104}
  {\path{doi:10.1002/prop.19860340104}}.

\bibitem{Ji:2017oey}
X.~Ji, J.-H. Zhang, Y.~Zhao, {Renormalization in Large Momentum Effective
  Theory of Parton Physics}, Phys. Rev. Lett. 120~(11) (2018) 112001.
\newblock \href {http://arxiv.org/abs/1706.08962} {\path{arXiv:1706.08962}},
  \href {http://dx.doi.org/10.1103/PhysRevLett.120.112001}
  {\path{doi:10.1103/PhysRevLett.120.112001}}.

\bibitem{Green:2017xeu}
J.~Green, K.~Jansen, F.~Steffens, {Nonperturbative renormalization of nonlocal
  quark bilinears for quasi-PDFs on the lattice using an auxiliary field},
  Phys. Rev. Lett. 121~(2) (2018) 022004.
\newblock \href {http://arxiv.org/abs/1707.07152} {\path{arXiv:1707.07152}},
  \href {http://dx.doi.org/10.1103/PhysRevLett.121.022004}
  {\path{doi:10.1103/PhysRevLett.121.022004}}.

\bibitem{Wang:2017eel}
W.~Wang, S.~Zhao, {On the power divergence in quasi gluon distribution function
  \ }\href {http://arxiv.org/abs/1712.09247} {\path{arXiv:1712.09247}}.

\bibitem{Alexandrou:2017huk}
C.~Alexandrou, K.~Cichy, M.~Constantinou, K.~Hadjiyiannakou, K.~Jansen,
  H.~Panagopoulos, F.~Steffens, {A complete non-perturbative renormalization
  prescription for quasi-PDFs}, Nucl. Phys. B923 (2017) 394--415.
\newblock \href {http://arxiv.org/abs/1706.00265} {\path{arXiv:1706.00265}},
  \href {http://dx.doi.org/10.1016/j.nuclphysb.2017.08.012}
  {\path{doi:10.1016/j.nuclphysb.2017.08.012}}.

\bibitem{Alexandrou:2017qpu}
C.~Alexandrou, K.~Cichy, M.~Constantinou, K.~Hadjiyiannakou, K.~Jansen,
  H.~Panagopoulos, A.~Scapellato, F.~Steffens, {Progress in computing parton
  distribution functions from the quasi-PDF approach}, EPJ Web of Conferences
  175 (Lattice 2017) (2018) 06021.
\newblock \href {http://arxiv.org/abs/1709.07513} {\path{arXiv:1709.07513}},
  \href {http://dx.doi.org/10.1051/epjconf/201817506021}
  {\path{doi:10.1051/epjconf/201817506021}}.

\bibitem{Alexandrou:2017dzj}
C.~Alexandrou, S.~Bacchio, K.~Cichy, M.~Constantinou, K.~Hadjiyiannakou,
  K.~Jansen, G.~Koutsou, A.~Scapellato, F.~Steffens, {Computation of parton
  distributions from the quasi-PDF approach at the physical point}, EPJ Web of
  Conferences 175 (Lattice 2017) (2018) 14008.
\newblock \href {http://arxiv.org/abs/1710.06408} {\path{arXiv:1710.06408}},
  \href {http://dx.doi.org/10.1051/epjconf/201817514008}
  {\path{doi:10.1051/epjconf/201817514008}}.

\bibitem{Chen:2017mzz}
J.-W. Chen, T.~Ishikawa, L.~Jin, H.-W. Lin, Y.-B. Yang, J.-H. Zhang, Y.~Zhao,
  {Parton distribution function with nonperturbative renormalization from
  lattice QCD}, Phys. Rev. D97~(1) (2018) 014505.
\newblock \href {http://arxiv.org/abs/1706.01295} {\path{arXiv:1706.01295}},
  \href {http://dx.doi.org/10.1103/PhysRevD.97.014505}
  {\path{doi:10.1103/PhysRevD.97.014505}}.

\bibitem{Lin:2017ani}
H.-W. Lin, J.-W. Chen, T.~Ishikawa, J.-H. Zhang, {Improved Parton Distribution
  Functions at Physical Pion Mass \ }\href {http://arxiv.org/abs/1708.05301}
  {\path{arXiv:1708.05301}}.

\bibitem{Stewart:2017tvs}
I.~W. Stewart, Y.~Zhao, {Matching the Quasi Parton Distribution in a Momentum
  Subtraction Scheme}, Phys. Rev. D97~(5) (2018) 054512.
\newblock \href {http://arxiv.org/abs/1709.04933} {\path{arXiv:1709.04933}},
  \href {http://dx.doi.org/10.1103/PhysRevD.97.054512}
  {\path{doi:10.1103/PhysRevD.97.054512}}.

\bibitem{Chen:2017mie}
J.-W. Chen, T.~Ishikawa, L.~Jin, H.-W. Lin, Y.-B. Yang, J.-H. Zhang, Y.~Zhao,
  {Operator classification for nonlocal quark bilinear on lattice \ }\href
  {http://arxiv.org/abs/1710.01089} {\path{arXiv:1710.01089}}.

\bibitem{Radyushkin:2017cyf}
A.~V. Radyushkin, {Quasi-parton distribution functions, momentum distributions,
  and pseudo-parton distribution functions}, Phys. Rev. D96~(3) (2017) 034025.
\newblock \href {http://arxiv.org/abs/1705.01488} {\path{arXiv:1705.01488}},
  \href {http://dx.doi.org/10.1103/PhysRevD.96.034025}
  {\path{doi:10.1103/PhysRevD.96.034025}}.

\bibitem{Orginos:2017kos}
K.~Orginos, A.~Radyushkin, J.~Karpie, S.~Zafeiropoulos, {Lattice QCD
  exploration of parton pseudo-distribution functions}, Phys. Rev. D96~(9)
  (2017) 094503.
\newblock \href {http://arxiv.org/abs/1706.05373} {\path{arXiv:1706.05373}},
  \href {http://dx.doi.org/10.1103/PhysRevD.96.094503}
  {\path{doi:10.1103/PhysRevD.96.094503}}.

\bibitem{Radyushkin:2018cvn}
A.~Radyushkin, {One-loop evolution of parton pseudo-distribution functions on
  the lattice \ }\href {http://arxiv.org/abs/1801.02427}
  {\path{arXiv:1801.02427}}.

\bibitem{Zhang:2018ggy}
J.-H. Zhang, J.-W. Chen, C.~Monahan, {Parton distribution functions from
  reduced Ioffe-time distributions}, Phys. Rev. D97~(7) (2018) 074508.
\newblock \href {http://arxiv.org/abs/1801.03023} {\path{arXiv:1801.03023}},
  \href {http://dx.doi.org/10.1103/PhysRevD.97.074508}
  {\path{doi:10.1103/PhysRevD.97.074508}}.

\bibitem{Chambers:2017dov}
A.~J. Chambers, R.~Horsley, Y.~Nakamura, H.~Perlt, P.~E.~L. Rakow,
  G.~Schierholz, A.~Schiller, K.~Somfleth, R.~D. Young, J.~M. Zanotti, {Nucleon
  Structure Functions from Operator Product Expansion on the Lattice}, Phys.
  Rev. Lett. 118~(24) (2017) 242001.
\newblock \href {http://arxiv.org/abs/1703.01153} {\path{arXiv:1703.01153}},
  \href {http://dx.doi.org/10.1103/PhysRevLett.118.242001}
  {\path{doi:10.1103/PhysRevLett.118.242001}}.

\bibitem{Ma:2017pxb}
Y.-Q. Ma, J.-W. Qiu, {Exploring Partonic Structure of Hadrons Using ab initio
  Lattice QCD Calculations}, Phys. Rev. Lett. 120~(2) (2018) 022003.
\newblock \href {http://arxiv.org/abs/1709.03018} {\path{arXiv:1709.03018}},
  \href {http://dx.doi.org/10.1103/PhysRevLett.120.022003}
  {\path{doi:10.1103/PhysRevLett.120.022003}}.

\bibitem{Chen:2017lnm}
J.-W. Chen, T.~Ishikawa, L.~Jin, H.-W. Lin, A.~Schäfer, Y.-B. Yang, J.-H.
  Zhang, Y.~Zhao, {Gaussian-weighted Parton Quasi-distribution \ }\href
  {http://arxiv.org/abs/1711.07858} {\path{arXiv:1711.07858}}.

\bibitem{Buras:1989xd}
A.~J. Buras, P.~H. Weisz, {QCD Nonleading Corrections to Weak Decays in
  Dimensional Regularization and 't Hooft-Veltman Schemes}, Nucl. Phys. B333
  (1990) 66--99.
\newblock \href {http://dx.doi.org/10.1016/0550-3213(90)90223-Z}
  {\path{doi:10.1016/0550-3213(90)90223-Z}}.

\bibitem{Patel:1992vu}
A.~Patel, S.~R. Sharpe, {Perturbative corrections for staggered fermion
  bilinears}, Nucl. Phys. B395 (1993) 701--732.
\newblock \href {http://arxiv.org/abs/hep-lat/9210039}
  {\path{arXiv:hep-lat/9210039}}, \href
  {http://dx.doi.org/10.1016/0550-3213(93)90054-S}
  {\path{doi:10.1016/0550-3213(93)90054-S}}.

\bibitem{Larin:1993tp}
S.~A. Larin, J.~A.~M. Vermaseren, {The Three loop QCD Beta function and
  anomalous dimensions}, Phys. Lett. B303 (1993) 334--336.
\newblock \href {http://arxiv.org/abs/hep-ph/9302208}
  {\path{arXiv:hep-ph/9302208}}, \href
  {http://dx.doi.org/10.1016/0370-2693(93)91441-O}
  {\path{doi:10.1016/0370-2693(93)91441-O}}.

\bibitem{Larin:1993tq}
S.~A. Larin, {The Renormalization of the axial anomaly in dimensional
  regularization}, Phys. Lett. B303 (1993) 113--118.
\newblock \href {http://arxiv.org/abs/hep-ph/9302240}
  {\path{arXiv:hep-ph/9302240}}, \href
  {http://dx.doi.org/10.1016/0370-2693(93)90053-K}
  {\path{doi:10.1016/0370-2693(93)90053-K}}.

\bibitem{Skouroupathis:2008mf}
A.~Skouroupathis, H.~Panagopoulos, {Two-loop renormalization of vector,
  axial-vector and tensor fermion bilinears on the lattice}, Phys. Rev. D79
  (2009) 094508.
\newblock \href {http://arxiv.org/abs/0811.4264} {\path{arXiv:0811.4264}},
  \href {http://dx.doi.org/10.1103/PhysRevD.79.094508}
  {\path{doi:10.1103/PhysRevD.79.094508}}.

\bibitem{Constantinou:2013pba}
M.~Constantinou, M.~Costa, H.~Panagopoulos, {Perturbative renormalization
  functions of local operators for staggered fermions with stout improvement},
  Phys. Rev. D88 (2013) 034504.
\newblock \href {http://arxiv.org/abs/1305.1870} {\path{arXiv:1305.1870}},
  \href {http://dx.doi.org/10.1103/PhysRevD.88.034504}
  {\path{doi:10.1103/PhysRevD.88.034504}}.

\bibitem{Boyle:2016wis}
P.~Boyle, L.~Del~Debbio, A.~Khamseh, {Massive momentum-subtraction scheme},
  Phys. Rev. D95~(5) (2017) 054505.
\newblock \href {http://arxiv.org/abs/1611.06908} {\path{arXiv:1611.06908}},
  \href {http://dx.doi.org/10.1103/PhysRevD.95.054505}
  {\path{doi:10.1103/PhysRevD.95.054505}}.

\bibitem{Gracey:2003yr}
J.~A. Gracey, {Three loop anomalous dimension of nonsinglet quark currents in
  the RI-prime scheme}, Nucl. Phys. B662 (2003) 247--278.
\newblock \href {http://arxiv.org/abs/hep-ph/0304113}
  {\path{arXiv:hep-ph/0304113}}, \href
  {http://dx.doi.org/10.1016/S0550-3213(03)00335-3}
  {\path{doi:10.1016/S0550-3213(03)00335-3}}.

\bibitem{Stefanis:1983ke}
N.~G. Stefanis, {Gauge-invariant quark two-point Green's function through
  connector insertion to $\mathcal{O} (\alpha_s)$}, Nuovo Cim. A83 (1984) 205.
\newblock \href {http://dx.doi.org/10.1007/BF02902597}
  {\path{doi:10.1007/BF02902597}}.

\bibitem{Chetyrkin:2003vi}
K.~G. Chetyrkin, A.~G. Grozin, {Three loop anomalous dimension of the heavy
  light quark current in HQET}, Nucl. Phys. B666 (2003) 289--302.
\newblock \href {http://arxiv.org/abs/hep-ph/0303113}
  {\path{arXiv:hep-ph/0303113}}, \href
  {http://dx.doi.org/10.1016/S0550-3213(03)00490-5}
  {\path{doi:10.1016/S0550-3213(03)00490-5}}.

\bibitem{Ji:1991pr}
X.-D. Ji, M.~J. Musolf, {Subleading logarithmic mass dependence in heavy meson
  form-factors}, Phys. Lett. B257 (1991) 409--413.
\newblock \href {http://dx.doi.org/10.1016/0370-2693(91)91916-J}
  {\path{doi:10.1016/0370-2693(91)91916-J}}.

\bibitem{Broadhurst:1991fz}
D.~J. Broadhurst, A.~G. Grozin, {Two loop renormalization of the effective
  field theory of a static quark}, Phys. Lett. B267 (1991) 105--110.
\newblock \href {http://arxiv.org/abs/hep-ph/9908362}
  {\path{arXiv:hep-ph/9908362}}, \href
  {http://dx.doi.org/10.1016/0370-2693(91)90532-U}
  {\path{doi:10.1016/0370-2693(91)90532-U}}.

\bibitem{tHooft:1973wag}
G.~'t~Hooft, M.~J.~G. Veltman, {Diagrammar}, Nato Sci. Ser. B 4 (1974)
  177--322.
\newblock \href {http://dx.doi.org/10.5170/CERN-1973-009}
  {\path{doi:10.5170/CERN-1973-009}}.

\end{thebibliography}

\end{document}